\documentclass[preprintnumbers,amssymb,amsmath,superscriptaddress,letterpaper,nofootinbib,twocolumn,floatfix,balancelastpage]{revtex4-1}[10pt]
\usepackage{graphicx,epstopdf,epsfig,bm,dcolumn,natbib,amsmath,amssymb,multirow}

\usepackage[
	colorlinks=true,
	citecolor=black,
	linkcolor=black,
	urlcolor=black,
	hypertexnames=false]{hyperref}

\graphicspath{{figures/}}

\newcommand{\beq}{\begin{equation}} \newcommand{\eeq}{\end{equation}}
\newcommand{\bea}{\begin{eqnarray}} \newcommand{\eea}{\end{eqnarray}}

\def\lsim{\mathrel{\raise.3ex\hbox{$<$\kern-.75em\lower1ex\hbox{$\sim$}}}}
\def\gsim{\mathrel{\raise.3ex\hbox{$>$\kern-.75em\lower1ex\hbox{$\sim$}}}}

\begin{document}


\hspace{12cm}
\preprint{FERMILAB-PUB-15-292-A}
\preprint{UCI-HEP-TR-2015-11}

\vspace{-1.7cm}

\title{Searching for MeV-Scale Gauge Bosons with IceCube}

\author{Anthony DiFranzo}
\affiliation{Center for Particle Astrophysics, Fermi National Accelerator Laboratory, Batavia, Illinois 60510, USA}
\affiliation{Department of Physics and Astronomy, University of California, Irvine, California 92697, USA}

\author{Dan Hooper}
\affiliation{Center for Particle Astrophysics, Fermi National Accelerator Laboratory, Batavia, IL 60510}
\affiliation{Department of Astronomy and Astrophysics, University of Chicago, Chicago, IL 60637}

\begin{abstract}

Light gauge bosons can lead to resonant interactions between high-energy astrophysical neutrinos and the cosmic neutrino background. We study this possibility in detail, considering the ability of IceCube to probe such scenarios. We find the most dramatic effects in models with a very light $Z'$ ($m_{Z'} \lsim 10$ MeV), which can induce a significant absorption feature at $E_{\nu} \sim$\, 5--10$\,{\rm TeV} \times (m_{Z'}/{\rm MeV})^2$. In the case of the inverted hierarchy and a small sum of neutrino masses, such a light $Z'$ can result in a broad and deep spectral feature at $\sim$\,0.1--10$\,{\rm PeV} \times (m_{Z'}/{\rm MeV})^2$. Current IceCube data already excludes this case for a $Z'$ lighter than a few MeV and couplings greater than $g\sim10^{-4}$. We emphasize that the ratio of neutrino flavors observed by IceCube can be used to further increase their sensitivity to $Z'$ models and to other exotic physics scenarios.

\end{abstract}

\maketitle

\section{Introduction}

In addition to providing a new window into the origin of the cosmic ray spectrum, the observation of astrophysical neutrinos allows us to probe fundamental physics. More specifically, IceCube's recent detection of high-energy astrophysical neutrinos enables us to study and constrain a range of phenomena at higher energies and over longer baselines than can currently be tested in laboratory environments. The flavor ratios of high-energy astrophysical neutrinos can be used to constrain a variety of new phenomena, including neutrino decay and Lorentz violation~\cite{Beacom:2002vi,Beacom:2003nh,Beacom:2003eu,Beacom:2003zg,Pagliaroli:2015rca,Bustamante:2015waa,Aeikens:2014yga,Hollander:2013im,Fu:2012zr,Bhattacharya:2010xj,Bustamante:2010nq,Bhattacharya:2009tx,Xing:2008fg,Hooper:2005jp,Hooper:2004xr,Arguelles:2015dca,Farzan:2014gza,Chatterjee:2013tza}. Alternatively, a few PeV neutrino interacting with a nucleon at rest has a center-of-mass energy of a few TeV, enabling high energy neutrino telescopes to constrain a range of TeV-scale physics scenarios, including TeV-scale gravity models~\cite{Illana:2014bda,AlvarezMuniz:2002ga,AlvarezMuniz:2001mk,Friess:2002cc,Han:2004kq}, and models featuring TeV-scale leptoquarks~\cite{Dutta:2015dka,Dey:2015eaa}.  These observations also make it possible to study interactions between high-energy neutrinos and the cosmic ray neutrino background (C$\nu$B). In particular, scenarios featuring a light gauge boson have received some recent attention within this context~\cite{Hooper:2007jr,Araki:2014ona,Ng:2014pca,Araki:2015dia,Kamada:2015era,Ibe:2014pja,Ioka:2014kca,Farzan:2015doa,Cherry:2014xra}. 

A new gauge boson with couplings to Standard Model neutrinos will induce a scattering resonance with the C$\nu$B at an energy given by:
\begin{eqnarray}
E^{\rm res}_{\nu} &\approx& \frac{m^2_{Z'}}{2m_{\nu}} 
\approx 1\,{\rm PeV} \times \bigg(\frac{m_{Z'}}{10 \, {\rm MeV}}\bigg)^2 \, \bigg(\frac{0.05 \,{\rm eV}}{m_{\nu}}\bigg).
\end{eqnarray}
For even very small couplings, such a resonance can lead to the efficient absorption of high-energy neutrinos over cosmological distances. 

In this paper, we revisit the possibility of using IceCube (or future high-energy neutrino telescopes) to search for the effects of an MeV-scale gauge boson on the high-energy cosmic neutrino spectrum. In doing so, we consider the impact on both the shape of the neutrino spectrum, as well as on the ratio of flavors that reach the Earth. In the window of parameter space that is capable of explaining the measured value of the muon's anomalous magnetic moment, significant effects can result from such a $Z'$. 
 
In the following two sections, we review IceCube's discovery of high-energy astrophysical neutrinos, and summarize the motivations for a model with an MeV-scale $Z'$ with couplings to Standard Model neutrinos. In Sec.~\ref{interactions}, we describe the interactions mediated by such a $Z'$ between high-energy neutrinos and the cosmic neutrino background. The impact of such interactions on the spectrum and the flavor ratios of the high-energy astrophysical neutrino flux is discussed in Sec.~\ref{results}. In Sec.~\ref{conclusion}, we summarize our results and conclusions.

\section{IceCube's Observation of High-Energy Astrophysical Neutrinos}

Recently, the IceCube Collaboration has reported the observation of a diffuse flux of high-energy extraterrestrial neutrinos, consisting of 37 neutrino candidate events with energies ranging from 30 TeV to 2 PeV~\cite{Aartsen:2013bka,Aartsen:2013jdh,Aartsen:2014gkd}.  Although the origin of these neutrinos is currently unknown, they appear to be approximately isotropically distributed across the sky~\cite{Aartsen:2014gkd}, suggesting an extragalactic origin (see, however, Refs.~\cite{Ahlers:2015moa,Neronov:2013lza,Ahlers:2013xia,Gupta:2013xfa,Lunardini:2013gva,Joshi:2013aua,Taylor:2014hya}). The spectrum of these particles is well fit by a power-law with an index of $\gamma=-2.6$~\cite{Aartsen:2015ivb}. 

Several features of this neutrino population are suggestive of a connection with the cosmic ray spectrum. In particular, the generation of the observed neutrino flux requires that $\sim$20\% of the PeV-EeV protons accelerated by cosmic ray sources undergo photo-meson interactions; a fraction that could easily be accommodated in realistic astrophysical environments~\cite{Cholis:2012kq}. Stated another way, the observed neutrino flux is below, but not very far below, what is known as the ``Waxman-Bahcall bound''~\cite{Waxman:1998yy,Bahcall:1999yr}. Furthermore, the numbers of showers and muon track events observed at IceCube is consistent with a flavor ratio of $\nu_e:\nu_{\mu}:\nu_{\tau}=1:1:1$ (although with large error bars)~\cite{Aartsen:2015ivb,Mena:2014sja}, consistent with that predicted from photo-meson interactions after accounting for oscillations. Several plausible classes of sources have been proposed for these neutrinos, including active galactic nuclei~\cite{Murase:2014foa,Dermer:2014vaa,Stecker:2013fxa,Murase:2014foa,Cholis:2012kq,Wang:2015woa}, starburst or star-forming galaxies~\cite{Liu:2013wia,He:2013cqa,Wang:2014jca,Tamborra:2014xia}, and low-luminosity gamma-ray bursts~\cite{Cholis:2012kq,Liu:2012pf,Murase:2013ffa,Xiao:2015gea}. In addition to these more conventional astrophysical source classes, exotic origins for IceCube's neutrinos have also been considered, such as the decays or annihilations of long-lived superheavy particles~\cite{Murase:2015gea,Kopp:2015bfa,Daikoku:2015vsa,Fong:2014bsa,Esmaili:2014rma,Rott:2014kfa,Ema:2014ufa,Bhattacharya:2014yha,Zavala:2014dla,Boucenna:2015tra}. In this study, we remain agnostic as to the specific origin of these neutrinos, assuming only that they originate from extragalactic sources.

\section{A Light $Z'$ with Couplings to Neutrinos}
\label{models}

New gauge bosons appear within many new physics scenarios~\cite{Langacker:2008yv}. For example, additional broken Abelian $U(1)$ gauge symmetries and the $Z'$ bosons that accompany them are predicted by many Grand Unified Theories (GUTs), including those based on the groups $SO(10)$ and $E_6$~\cite{London:1986dk,Hewett:1988xc}. New massive gauge bosons also appear within the context of many string inspired models~\cite{Braun:2005bw,Cleaver:1998gc,Coriano:2007ba,Faraggi:1990ita,Giedt:2000bi,Lebedev:2007hv,Anastasopoulos:2006cz,Faraggi:1991mu,Cvetic:2001nr}, little Higgs theories~\cite{ArkaniHamed:2001is,ArkaniHamed:2002qx,Han:2003wu,Perelstein:2005ka}, dynamical symmetry breaking scenarios~\cite{Hill:2002ap,Chivukula:2003wj,Chivukula:2002ry}, models with extra spatial dimensions~\cite{Agashe:2003zs,Agashe:2007ki,Carena:2003fx,Hewett:2002fe}, and many other proposed extensions of the Standard Model~\cite{ArkaniHamed:2001nc,Cvetic:1997ky,Langacker:1999hs}. 

In this paper, we are primarily interested in light gauge bosons ($m_{Z'} < 1$ GeV) with nonzero couplings to Standard Model neutrinos. Constraints on such a particle's couplings to electron neutrinos are quite stringent, however, motivating us to focus on models in which the $Z'$ couples only to 2nd and/or 3rd generation leptons. 

If the couplings of a $Z'$ are assigned arbitrarily, anomalies are generally introduced, violating the principle of gauge invariance. To construct a self-consistent theory, care must be taken to ensure that all such anomalies cancel. A well-known example of an anomaly-free $Z'$ is that arising from the gauge group $U(1)_{\mu-\tau}$. This is the only anomaly-free $U(1)$ group with nonzero charge assignments to Standard Model neutrinos that can lead to an experimentally viable MeV-scale $Z'$ without requiring the addition of any exotic fermions. A $U(1)$ group charged under only muon or tau number is also a possibility, although new chiral fermions must be introduced in these cases, charged under both $SU(2)_W$ and $U(1)_Y$, as well as under the new $U(1)_{\mu}$ or $U(1)_{\tau}$ group. 

A light $Z'$ with couplings to the muon can be motivated by the measurement of the muon's anomalous magnetic moment, which currently differs from the value predicted by the Standard Model with a significance of approximately 3.6$\sigma$~\cite{Beringer:1900zz}. With efforts currently underway to improve this measurement~\cite{Carey:2009zzb,Kronfeld:2013uoa} and to reduce the related theoretical uncertainties~\cite{Aubin:2012me,Aubin:2013daa,Aubin:2013yba,Blum:2013xva,Golterman:2013vna,Nyffeler:2013lia}, it should become clear within the next several years whether or not this is an authentic sign of new physics.  Among other possibilities (see, for example, Refs.~\cite{Endo:2013lva,Ibe:2013oha,Davoudiasl:2014kua,Agrawal:2014ufa,Ajaib:2014ana,McKeen:2009ny}), an MeV-scale $Z'$ with small couplings to the muon could plausibly account for this measurement.  

A $Z'$ with a vector coupling to muons, $g_{\mu}$, leads to the following contribution to the muon's magnetic moment~\cite{Jegerlehner:2009ry,Queiroz:2014zfa}:
\begin{equation}
\Delta a_{\mu} = \frac{g^2_{\mu}  m^2_{\mu}}{4 \pi^2 m^2_{Z'}} \int^1_0 dx \frac{x^2(1-x)}{1-x+(m^2_{\mu}/m^2_{Z'})x^2}.
\end{equation}
For $m_{Z'} \ll m_{\mu}$, the measured value can be accommodated for $g_{\mu} \sim (3-6) \times 10^{-4}$, whereas for $m_{Z'} \simeq $1 GeV, couplings an order of magnitude larger are required. Although this parameter space is in conflict with measurements of muon pair production in muon neutrino-nucleus scattering for $m_{Z'} \gsim 500$ MeV~\cite{Altmannshofer:2014pba,Araki:2014ona}, lower values of $m_{Z'}$ remain viable (below 1 MeV, constraints from big bang nuclosynthesis and the cosmic microwave background can also be relevant~\cite{Ahlgren:2013wba,Cyr-Racine:2013jua,Archidiacono:2013dua,Ng:2014pca}).

As stated above, a $Z'$ resulting from the $U(1)_{\mu}$ or $U(1)_{\tau}$ groups requires the introduction of new chiral fermions charged under $SU(2)_W$ and $U(1)_Y$ in order to cancel anomalies. Furthermore, the masses of these exotics are bounded by the requirement of perturbativity, which requires~\cite{Dobrescu:2014fca}:
\begin{equation}
m_{\rm exotic} \lsim 108 \, {\rm GeV} \times \bigg(\frac{m_{Z'}}{10 \, {\rm MeV}}\bigg) \bigg(\frac{0.0005}{g_{Z'}}\bigg) \bigg(\frac{1}{z_{\varphi}}\bigg),
\end{equation}
where $z_{\varphi}$ is the charge assignment for the scalar whose VEV breaks the $U(1)$ responsible for the $Z'$. From this, we learn that the required exotics must be rather light, and will be subject to constraints from accelerators.

\section{High Energy Neutrino Interactions with the Cosmic Neutrino Background}
\label{interactions}

In generality, a $Z'$ boson will couple to Standard Model neutrinos through the following interaction:
\begin{eqnarray}
\mathcal L_{Z'\nu} &= g_{Z'} Q_{\alpha \beta} Z'_{\mu} \overline\nu_{\alpha} \gamma^{\mu} P_L \nu_{\beta}\\
   &= g_{Z'} Q'_{i j} Z'_{\mu} \overline\nu_{i} \gamma^{\mu} P_L \nu_{j},
\end{eqnarray}
where neutrinos with Greek (Latin) indices refer to the flavor (mass) basis. The $U(1)'$ gauge coupling is $g_{Z'}$ and the charge assignments are contained in the matrix, $Q_{\alpha \beta}$. For example, the charge matrix for $U(1)_{\mu-\tau}$ is given by $Q_{\alpha \beta}=\text{diag}(0,1,-1)$. In the mass basis, the charge matrix is represented as $Q'_{ij} = U^{\dagger}_{\alpha i} Q_{\alpha \beta} U_{\beta j}$. Where $U_{\alpha i}$ is the Pontecorvo--Maki--Nakagawa--Sakata (PMNS) matrix which rotates from the mass to the flavor basis. Here we have assumed that the charged leptons do not mix, such that the neutrino mixing matrix is entirely determined by the PMNS matrix. While this is a convention in the Standard Model, deviating from this assumption does lead to observable effects in models with gauged lepton numbers \cite{Heeck:2011wj}.

This interaction results in $\nu_i-\nu_j$ scattering, with the following cross section:
\begin{equation}
\sigma(\nu_i \nu_j\rightarrow \nu\nu)= \sum_{k\ell}Q'^2_{k\ell} \frac{g_{Z'}^4 Q'^2_{ij}}{3\pi} \frac{s}{(s-m_{Z'}^2)^2+m_{Z'}^2\Gamma_{Z'}^2},
\label{eq:xsec}
\end{equation}
where the sum is over the final neutrino states, $k$ and $\ell$. For example, $\sum_{k\ell}Q'^2_{k\ell}=2$ for the $U(1)_{\mu-\tau}$ case. The $t$-channel contribution to the cross section is ignored, as it is highly suppressed relative to the $s$-channel resonance.  The width of the $Z'$ into neutrinos is given by $\Gamma_{Z'}=\sum_{ij}Q'^2_{ij} g_{Z'}^2 m_{Z'}/24\pi$. In the parameter space of interest to IceCube, the mass of the $Z'$ is less than $2m_{\mu}$, allowing us to safely neglect decays to muons or taus.

To calculate the spectrum of neutrinos at Earth, we solve the following coupled set of integro-differential equations \cite{Ng:2014pca,Blum:2014ewa,Araki:2015dia}:
\begin{eqnarray}
\label{eq:difeq}
&-&(1+z) \frac{H(z)}{c} ~ \frac{d\widetilde{n}_i}{dz} = J_i(E_0,z) \nonumber \\
&-&\widetilde{n}_i \sum_j \left<n_{\nu j}(z)~\sigma_{ij}(E_0,z)\right> \nonumber\\
 &+& P_i\int_{E_0}^\infty dE' \sum_{j,k} \widetilde{n}_k \left<n_{\nu j}(z)~\frac{d\sigma_{kj}}{dE_0}(E',z)\right>,
\end{eqnarray}
where
\begin{align*}
\widetilde{n}_i &\equiv \frac{dN_i}{dE}(E_0,z),\\
P_i &\equiv \sum_\ell Br(Z' \rightarrow \nu_\ell \nu_i).
\end{align*}
Here, $H(z)$ is the Hubble parameter as a function of redshift, $E_0$ is the neutrino energy as measured at Earth, and $n_{\nu}(z)$ is the proper number density of neutrinos in the $C\nu B$. The first term on the right-hand side accounts for the source spectral and density evolution through cosmology. The second term accounts for neutrinos of state $i$, scattering off a thermal distribution of $C\nu B$ neutrinos of state $j$, thereby attenuating the neutrino flux. The last term accounts for the regeneration of the scattering products from the process described in the second term. Here a neutrino of state $k$, with energy $E'$, scatters with the $C\nu B$ to produce two neutrinos, one of which is of state $i$ with energy $E_0$. Since this process is $s$-channel, the differential cross section can be broken into the total cross section and a distribution function in the outgoing neutrino energy space:
\begin{eqnarray}
\frac{d\sigma_{kj}}{dE_0}(E',z) = \sigma_{kj}(E',z) f(E',E_0),
\end{eqnarray}
where
\begin{eqnarray}
f(E',E_0) = \frac{3}{E'}\bigg[\left(\frac{E_0}{E'}\right)^2 \!+ \left(1-\frac{E_0}{E'} \right)^2\bigg] \Theta(E'-E_0). \nonumber
\end{eqnarray}
Note that this differential cross section accounts for both outgoing neutrinos, such that $\int dE_0 f(E',E_0) = 2$. Expressing the differential cross section in this way simplifies the numerical implementation of Eqn.~\ref{eq:difeq}.

The thermal averaging in the second term of Eqn.~\ref{eq:difeq}, and analogously in the third term, is given by:
\begin{eqnarray}
\left<n_{\nu j}(z) \sigma_{i j} \right> \equiv \int \frac{d^3\mathbf{p}}{(2\pi)^3} \frac{\sigma_{i j}(E_0,z,\mathbf{p})}{e^{|\mathbf{p}|/T_0(1+z)}+1} \nonumber \\
\xrightarrow{~m_j\gg T~} n_{\nu j}(z) ~\sigma_{ij}(E_0,z),
\end{eqnarray}
where the momentum dependence of the scattering cross section can be accounted for by evaluating Eqn.~\ref{eq:xsec} with $s=2E_0(1+z)\big(\sqrt{\smash[b]{m_j^2+p^2}}-|p|\cos\theta\big)$. In the limit that the neutrinos are much heavier than the effective temperature, the usual form is recovered. We take the effective temperature of the $C\nu B$ at $z=0$ to be $T_0=1.7\times 10^{-4}$ eV.

While an absorption resonance is predicted for each neutrino mass eigenstate, very light neutrinos (with masses comparable to or less than the effective temperature the thermal distribution of the $C\nu B$) will lead to broad spectral features. In scenarios in which the lightest neutrino is nearly massless, the energy of the resonance is determined by the energy (and therefore the temperature) of the neutrino, rather than by its mass. One then expects scattering to be most relevant at $E_{\nu} \sim m_{Z'}^2/2T_0$.\footnote{More precisely, one expects the cross section to peak at $E_{\nu} \sim m_{Z'}^2/2\!\left<p\right>$, with $\left<p\right>=\frac{7\pi^4 T}{180 \zeta(3)} \approx 3.15\,T$.} 

Throughout this paper, we adopt values for the neutrino mass splittings and mixing angles as presented in Ref.~\cite{Forero:2014bxa}, with the exception of the CP violating angle which we take to be $\delta{=}0$ (which is well within $2\sigma$ of the central value). For the cosmological parameters $H_0$, $\Omega_m$, and $\Omega_\lambda$, and for the upper limit on the sum of the neutrino masses, we adopt the values presented by the Planck Collaboration~\cite{Ade:2015xua}.


\section{The Impact of a Light $Z'$ on the Spectrum and Flavor Ratios of the High-Energy Neutrino Flux}
\label{results}

\subsection{Main Results}

In this section, we present the results of our calculations. As our primary model of interest, we consider a $Z'$ associated with the gauge group $U(1)_{\mu-\tau}$. For each value of $m_{Z'}$, the coupling $g_{Z'}$ is chosen to match the measured value of the muon's magnetic moment, $\Delta a_{\mu}$:
\begin{equation}
g_{Z'} \approx \begin{cases} 4\times10^{-4} & \text{for } m_{Z'}=1 \text{ MeV}\\
		5\times10^{-4} & \text{for } m_{Z'}=10 \text{ MeV}\\
		8\times10^{-4} & \text{for } m_{Z'}=100 \text{ MeV}.
\end{cases}
\end{equation}

\begin{figure*}
	\includegraphics[width=0.49\textwidth]{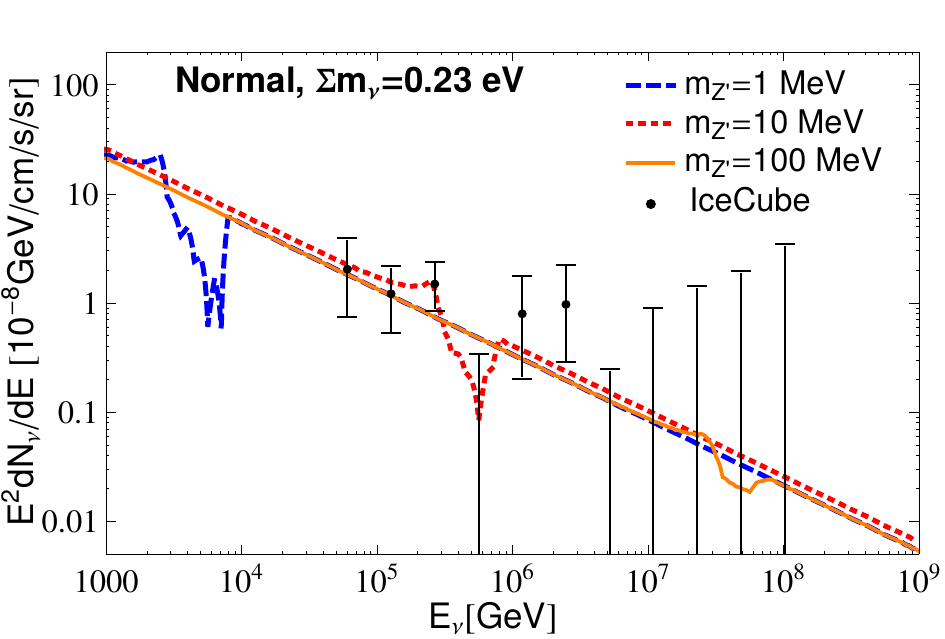}
\includegraphics[width=0.49\textwidth]{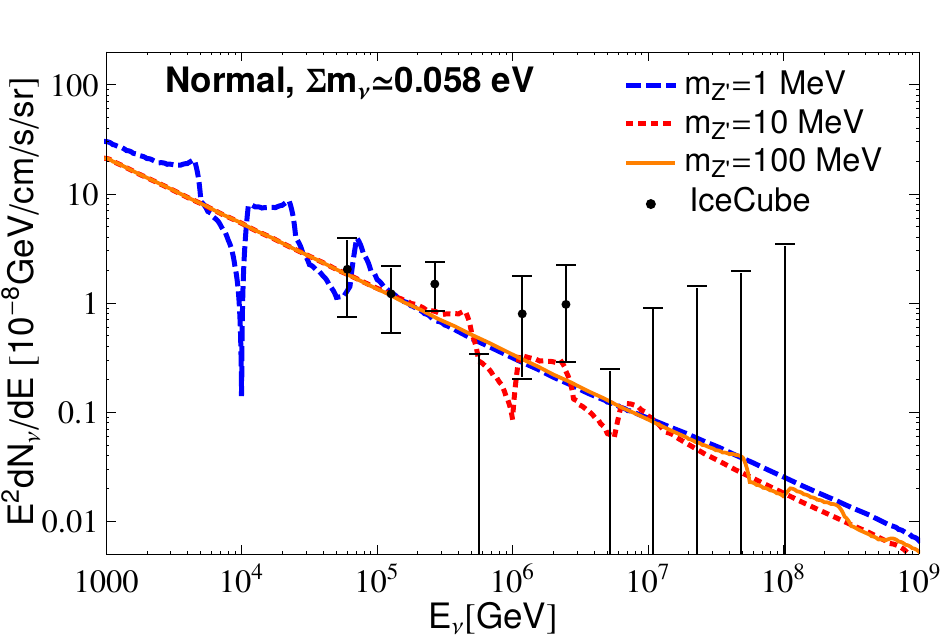}\\
	\includegraphics[width=0.49\textwidth]{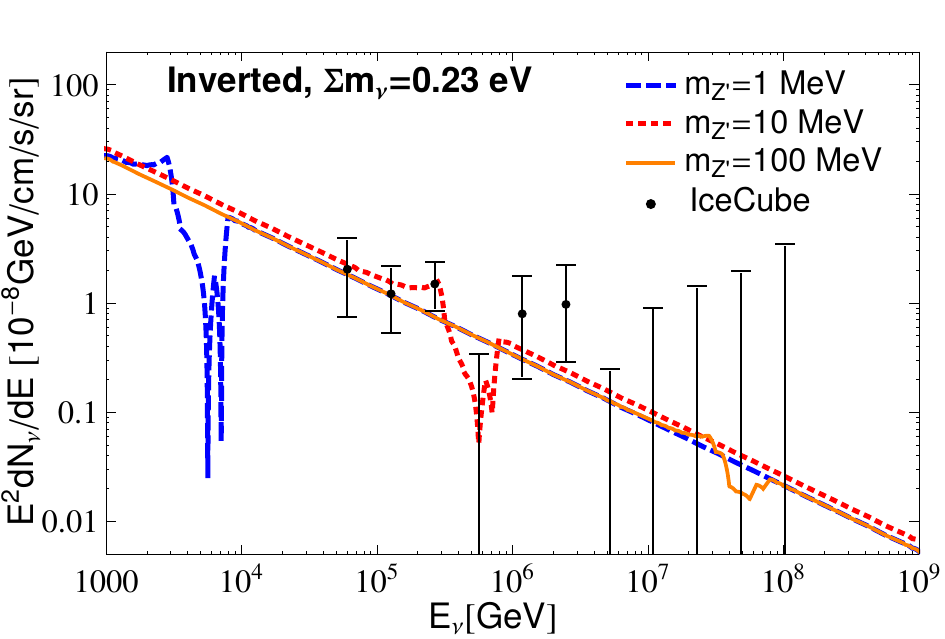}
		\includegraphics[width=0.49\textwidth]{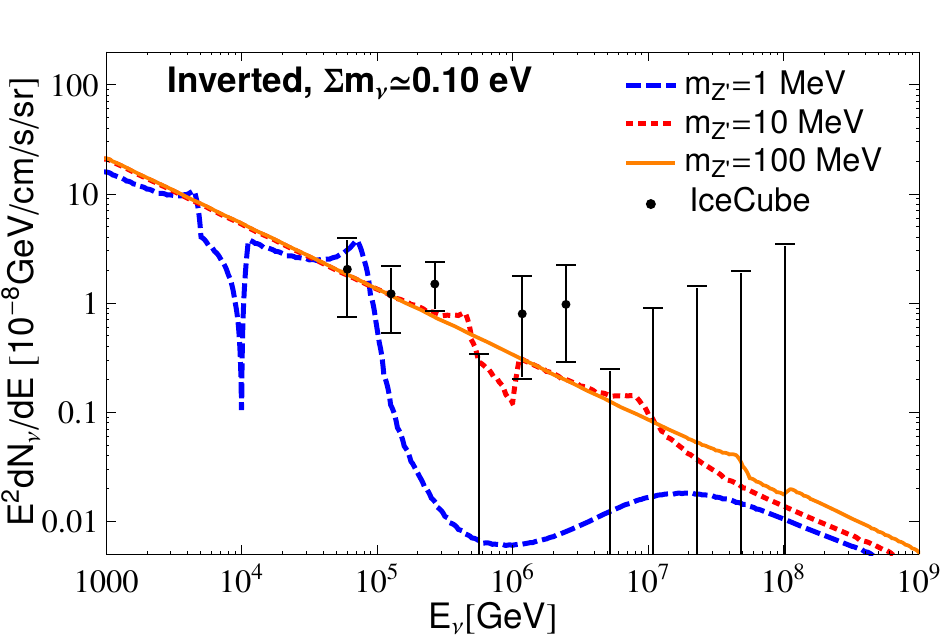} 
	\caption{The spectrum of neutrinos at Earth, after including the effects of a $Z'$ associated with the gauge group $U(1)_{\mu-\tau}$.  The coupling of the $Z'$ has been chosen in each case to accommodate the measured value of the muon's anomalous magnetic moment ($g_{Z'} = 4\times10^{-4}$, $5\times10^{-4}$ and $8\times10^{-4}$ for $m_{Z'}=1$, 10 and 100 MeV, respectively). Here, we have assumed a population of sources at $z=1$ which inject neutrinos with a power-law spectrum of index of -2.6, and with an initial flavor ratio of $\nu_e:\nu_{\mu}:\nu_{\tau}=1:2:0$ (as predicted from pion decay). We show results for the normal and inverted hierarchies, and for maximal and minimal values of the sum of the neutrino masses.}
\label{fig:IH023}
\end{figure*}

For simplicity, we first consider the case in which all of the neutrino sources are located at a redshift of $z=1$, and which emit a spectra characterized by a power-law of index $\gamma=-2.6$:
\begin{equation}
J(E_{\nu},z) =  A \, E_{\nu}^{\gamma} \, \delta(z-1),
\end{equation}
where the normalization, $A$, is chosen to fit the IceCube data. We adopt an initial flavor ratio of $\nu_e:\nu_{\mu}:\nu_{\tau}=1:2:0$ (as predicted from pion decay), which rapidly evolves to approximately $1:1:1$ via oscillations.

\begin{figure*}
	\includegraphics[width=0.32\textwidth]{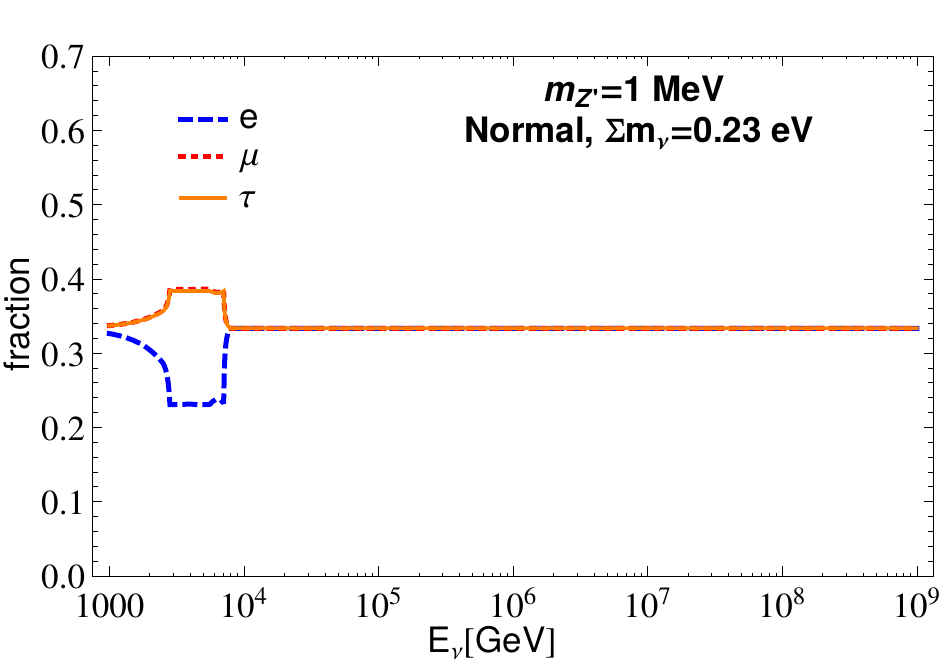}	
	\includegraphics[width=0.32\textwidth]{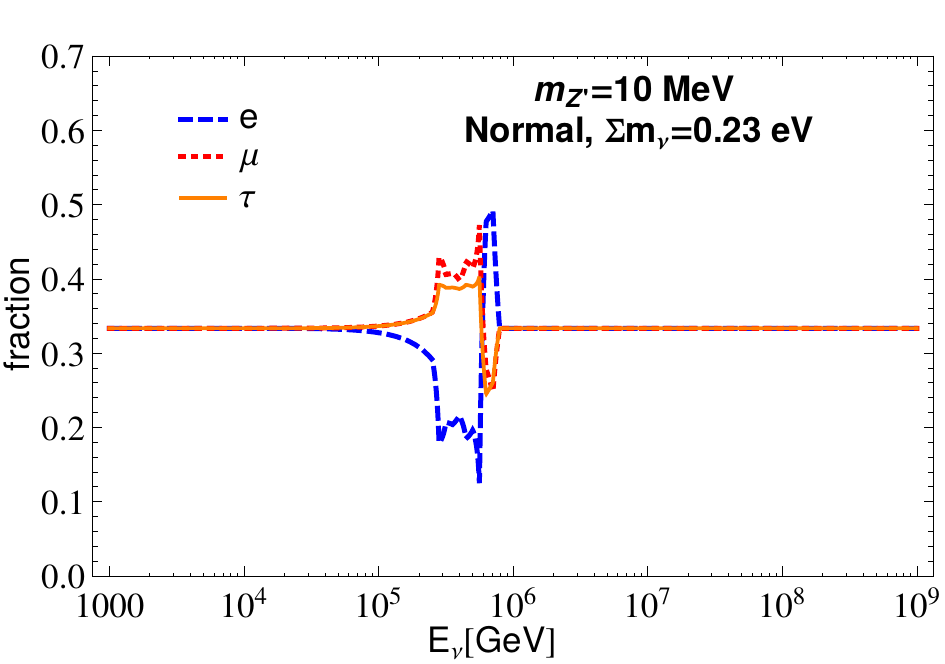}
	\includegraphics[width=0.32\textwidth]{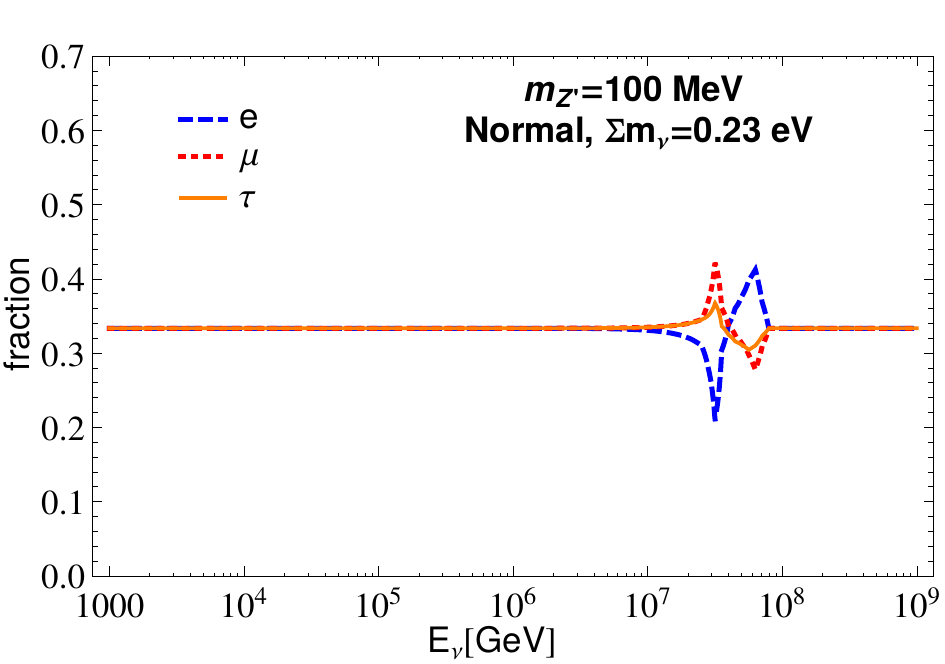}\\
	\includegraphics[width=0.32\textwidth]{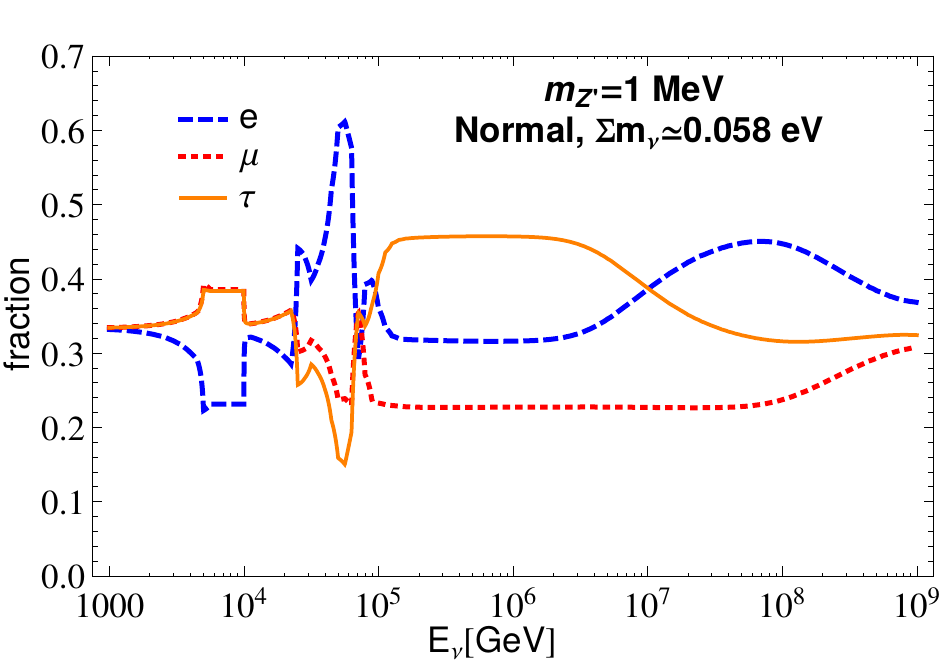}
	\includegraphics[width=0.32\textwidth]{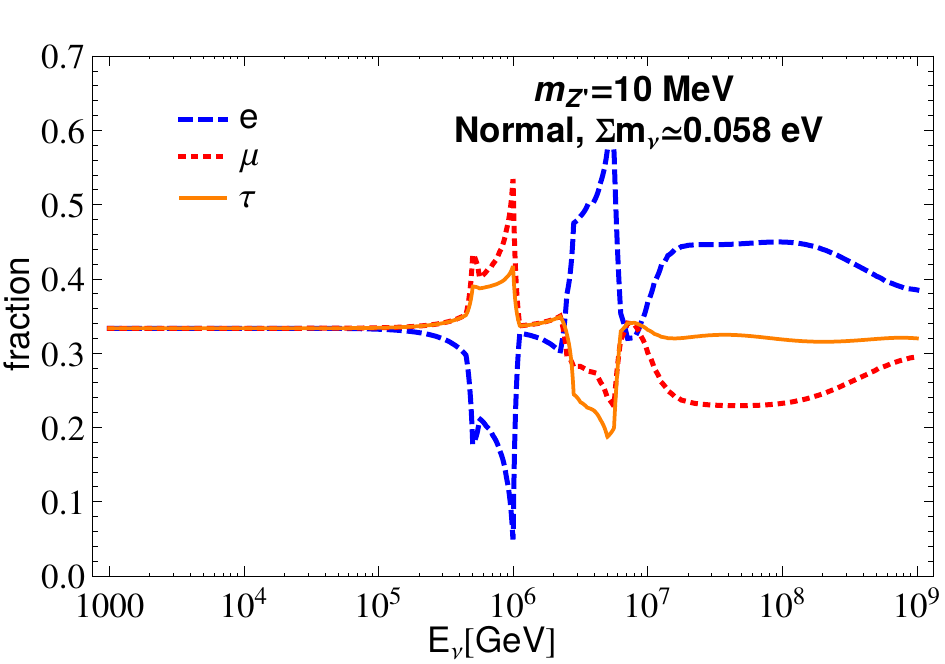}
	\includegraphics[width=0.32\textwidth]{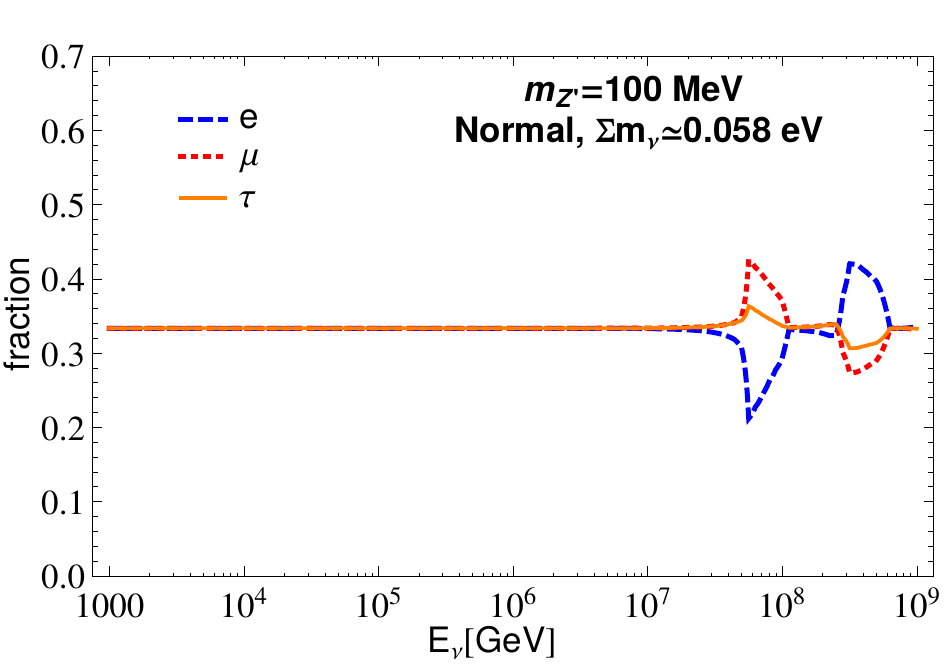}\\
	\includegraphics[width=0.32\textwidth]{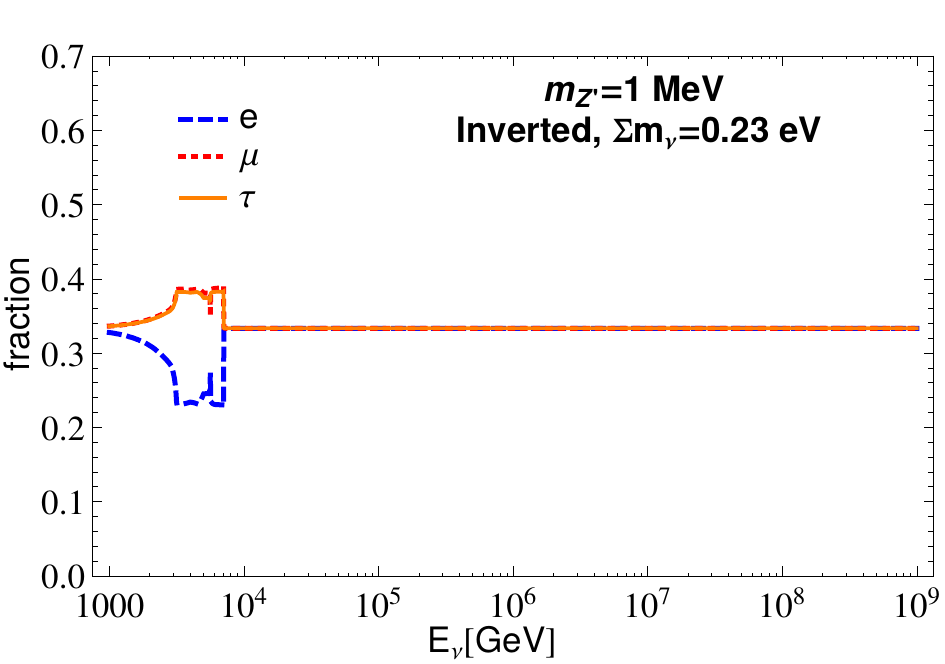}
	\includegraphics[width=0.32\textwidth]{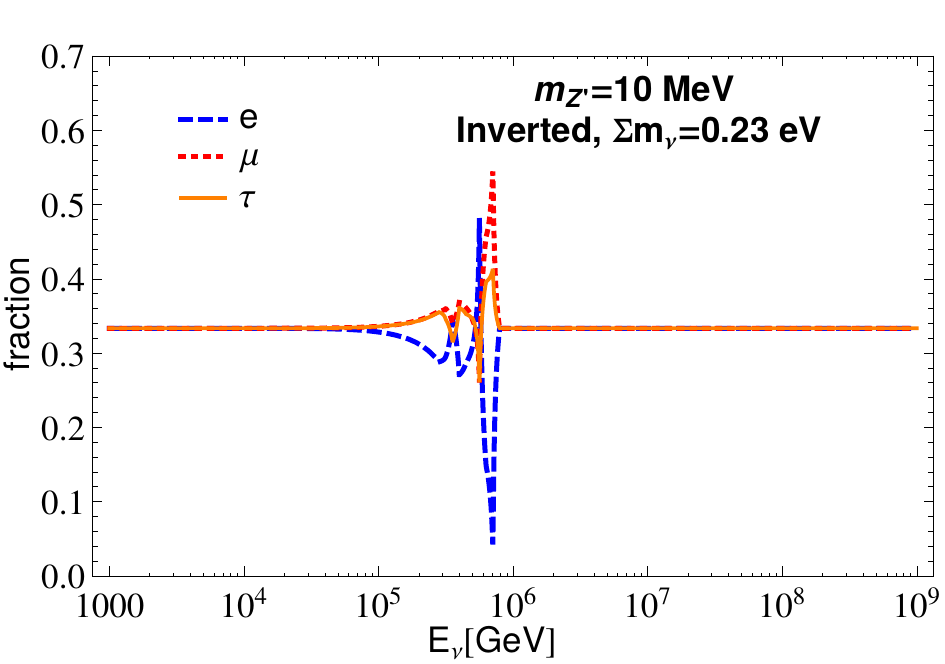}
\includegraphics[width=0.32\textwidth]{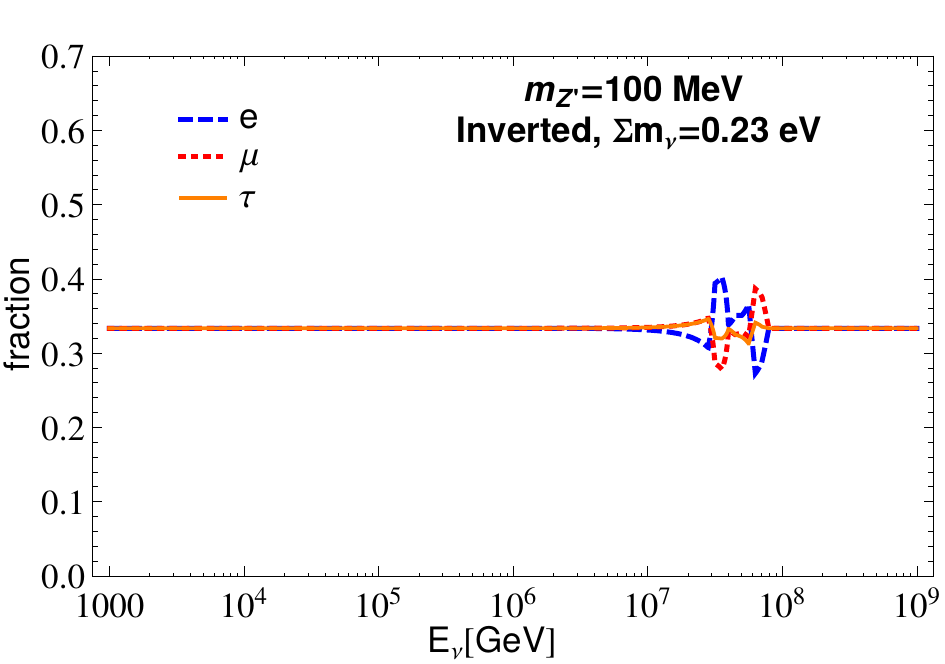}\\
\includegraphics[width=0.32\textwidth]{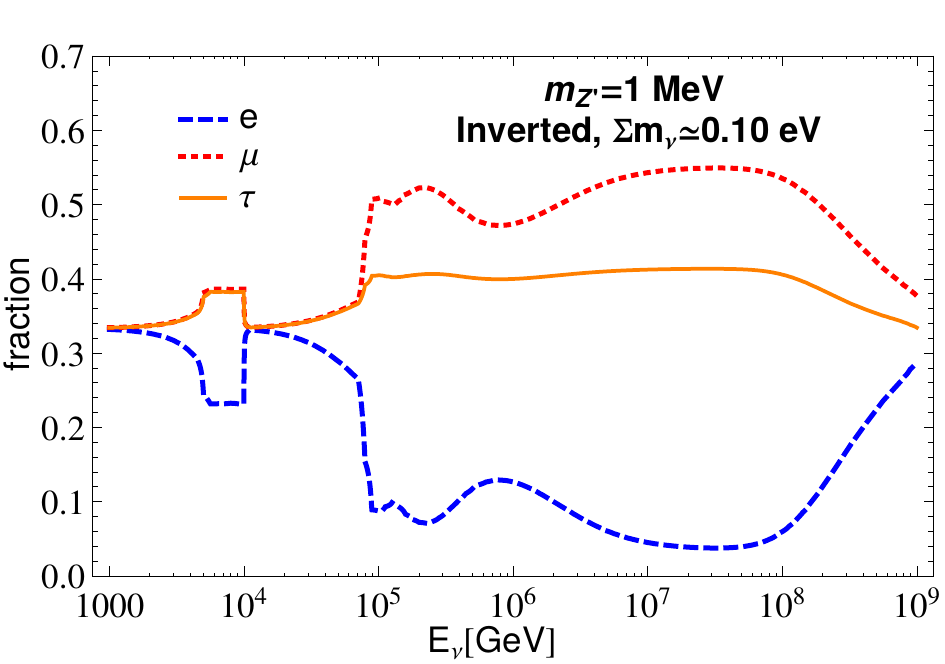}
		\includegraphics[width=0.32\textwidth]{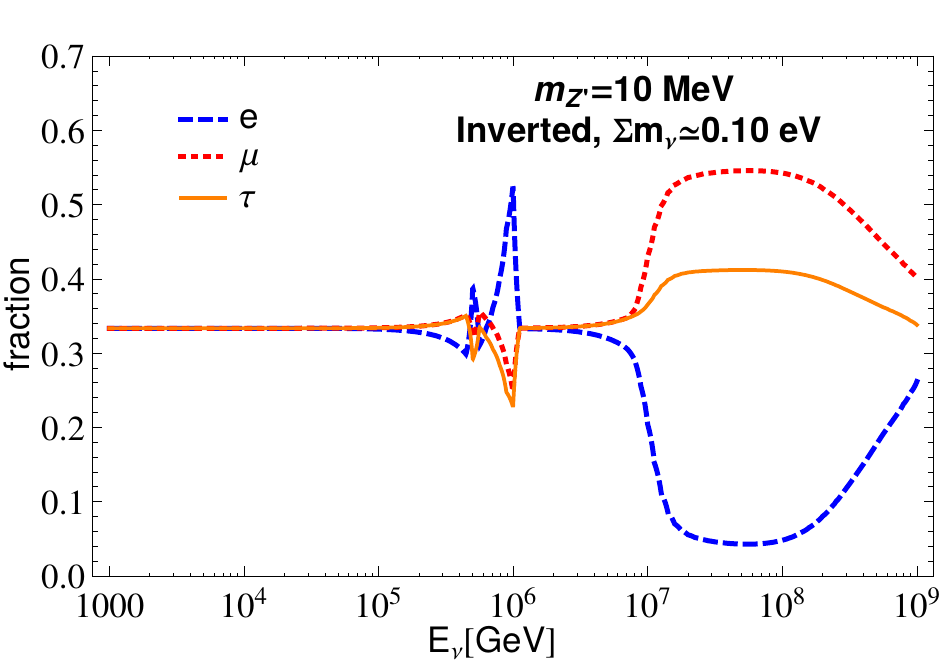}
	\includegraphics[width=0.32\textwidth]{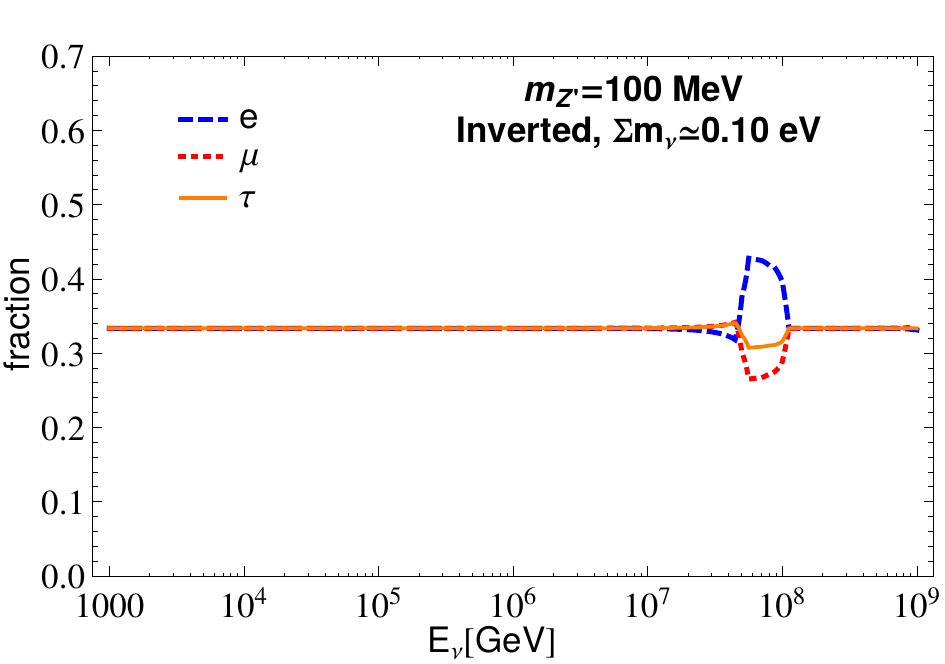}
	\caption{The fractions of neutrinos of each flavor at Earth, including the effects of a $Z'$. The models and other assumptions are the same as adopted in Fig.~\ref{fig:IH023}.}
\label{fig:flavor}
\end{figure*}

In Fig.~\ref{fig:IH023}, we plot the neutrino spectrum at Earth, including the effects of a $Z'$. Results are shown for three choices of $m_{Z'}$, and for both normal and inverted hierarchies, as well as maximal and minimal values for the sum of the neutrino masses. For each hierarchy, we take $\sum m_\nu {=} 0.23$ eV to be the maximal value allowed by cosmological constraints~\cite{Ade:2015xua}. For the minimal sum of masses, we adopt 0.058 eV and 0.10 eV for the normal and inverted hierarchies, respectively. 

In each case, the neutrino spectrum is altered by the interactions of the $Z'$, although in ways that vary considerably depending on the mass of the $Z'$ and on the hierarchy and the masses of the neutrinos. In the case of the normal hierarchy with $\sum m_\nu {=} 0.23$ eV, for example, an absorption feature appears at $E_{\nu} \simeq 5\,{\rm TeV} \times (m_{Z'}/{\rm MeV})^2$. This feature results from the scattering with all three neutrino mass eigenstates, and the individual resonances cannot be easily distinguished. In the normal hierarchy with the minimal sum of masses, two features are visible: one at $E_{\nu} \simeq 10 \,\,{\rm TeV} \times (m_{Z'}/{\rm MeV})^2$ from scattering with the heaviest mass eigenstate, and another at $E_{\nu} \simeq 50 \,\,{\rm TeV} \times (m_{Z'}/{\rm MeV})^2$ resulting from the combination of the two lighter mass eigenstates.  

The results are somewhat different in the case of the inverted hierarchy. As found for the normal hierarchy, an absorption feature appears at $E_{\nu} \simeq 5-10\,\,{\rm TeV} \times (m_{Z'}/{\rm MeV})^2$. In the case of $\sum m_\nu {=} 0.23$ eV, this is the collective consequence of all three mass eigenstates. For $\sum m_\nu {=} 0.10$ eV, however, this feature is induced only through scattering with the heaviest two eigenstates. In this case, the lightest neutrino leads instead to a very broad and potentially deep spectral feature, covering a range of energies between $E_{\nu} \sim\,$\,0.1--10$\,\,{\rm PeV} \times (m_{Z'}/{\rm MeV})^2$. This broad absorption feature is most clearly visible in the case of $m_{Z'}=1$ MeV. In this case, the spectrum reported by the IceCube Collaboration (shown as error bars) is incompatible with a $Z'$ lighter than a few MeV. Note that this feature does not appear in the case of the normal hierarchy because the lightest neutrino is largely of $\nu_e$ flavor, and thus does not couple to the $Z'$ under consideration. In contrast, the lightest mass eigenstate in the inverted hierarchy is primarily composed of $\nu_{\mu}$ and $\nu_{\tau}$.


In addition to the impact on the high-energy neutrino spectrum, a light $Z'$ can alter the ratio of neutrino flavors that reach Earth. In Fig.~\ref{fig:flavor}, we plot these ratios for the same range of scenarios considered in Fig.~\ref{fig:IH023}. Similar to the spectrum, the most significant effects are seen in the case of the inverted hierarchy with a sum of neutrino masses near the minimal value. In this case, the relative flux of electron (muon) neutrinos is suppressed (enhanced) over a wide range of energies, especially for the case of $m_{Z'} \lsim 10$ MeV. Such an extreme departure from astrophysical expectations could plausibly be tested in the future by IceCube.

Constraints on the flavor ratios of IceCube's neutrinos have been placed by comparing the distribution of muon track events (generated in charged current interactions of $\nu_{\mu}$ and $\bar{\nu}_{\mu}$) to the distribution of showers (generated most efficiently in charged current interactions of $\nu_{e}$, $\bar{\nu}_{e}$, $\nu_{\tau}$ and $\bar{\nu}_{\tau}$, as well as in neutral current interactions of all flavors). The results of this comparison have thus far been compatible with a ratio of $\nu_e:\nu_{\mu}:\nu_{\tau}=1:1:1$, although with large error bars~\cite{Aartsen:2015ivb}. More specifically, extreme ratios of $\nu_e:\nu_{\mu}:\nu_{\tau}=0:1:0$ and $\nu_e:\nu_{\mu}:\nu_{\tau}=1:0:0$ have been excluded at the level of 3.3$\sigma$ and 2.3$\sigma$ significance, respectively. 

Future flavor ratio measurements by IceCube are expected to be strengthened by improvements in veto techniques, and by searches for events unique to tau neutrinos.\footnote{At energies above a few PeV, tau neutrinos and antineutrinos can generate a tau lepton that travels an observable distance before decaying; the mean distance traveled is $R_{\tau} = E_{\tau} c \tau_{\tau}/m_{\tau} \simeq 50$\,m $\times E_{\tau}/(1\,{\rm PeV})$. As a result, very high-energy tau neutrinos can yield events with two showers (double bang events)~\cite{Learned:1994wg,Athar:2000rx}, as well as events with one observed shower followed by or preceded by a tau-induced track (lollipop events)~\cite{Beacom:2003nh}.} Showers generated via the Glashow resonance (at $E_{\bar{\nu}_e} = m^2_W/2 m_e \approx 6.3$ TeV) also provide an opportunity to constrain the electron antineutrino fraction at very high energies.  

\subsection{Alternative Gauge Groups}

\begin{figure}
	\includegraphics[width=0.5\textwidth]{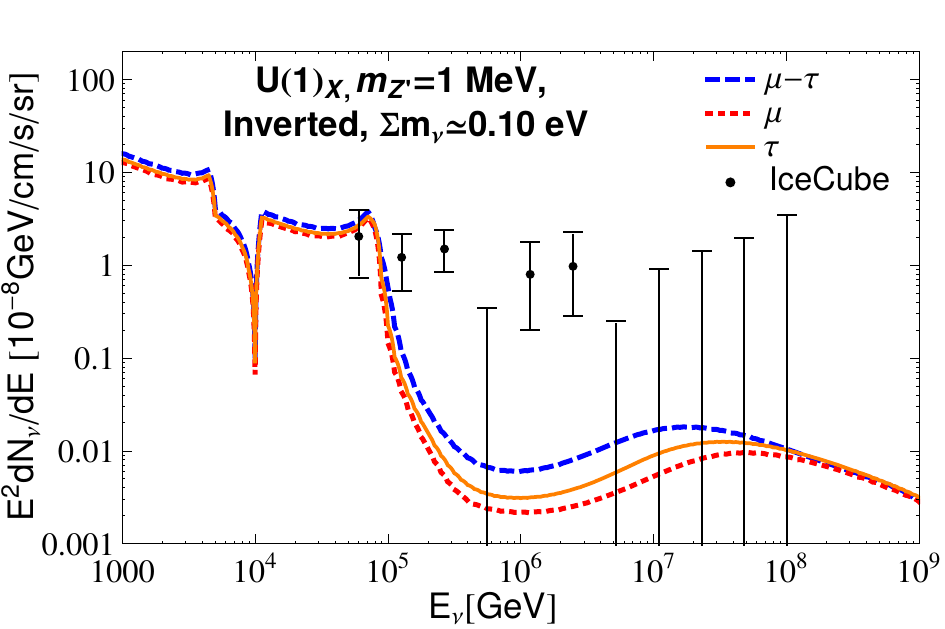}
\caption{A comparison of the neutrino spectra predicted for a $Z'$ associated with the $U(1)_{\mu-\tau}$ (blue), $U(1)_{\mu}$ (red), and $U(1)_{\tau}$ gauge groups. Results are shown for the case of an inverted hierarchy, $\sum m_{\nu}{\approx}0.10$ eV, and $m_{Z'}{=}1$ MeV.}
\label{fig:u1x_spec}
\end{figure}

\begin{figure}
	\includegraphics[width=0.5\textwidth]{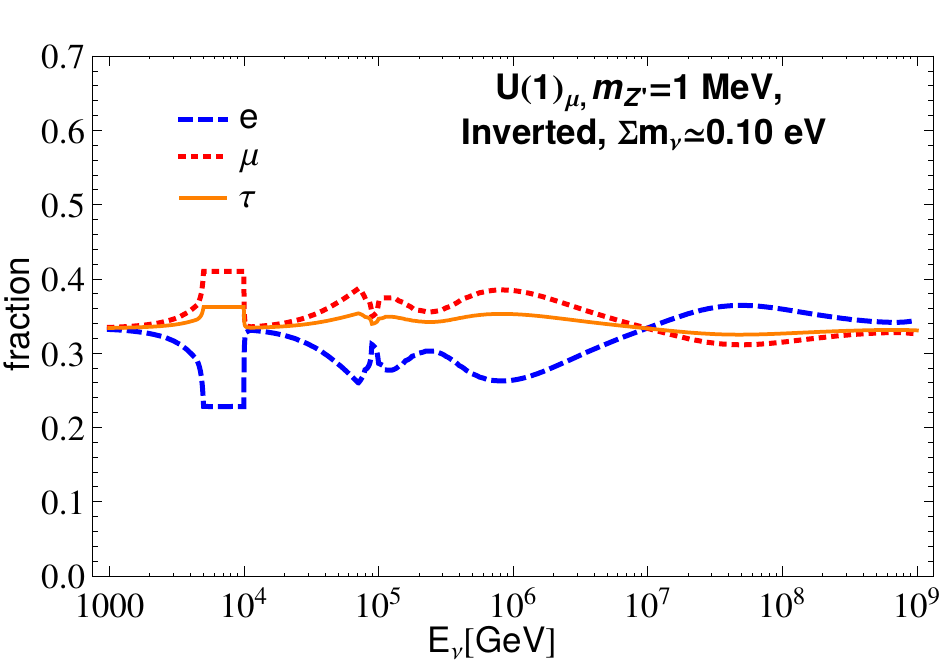}
		\includegraphics[width=0.5\textwidth]{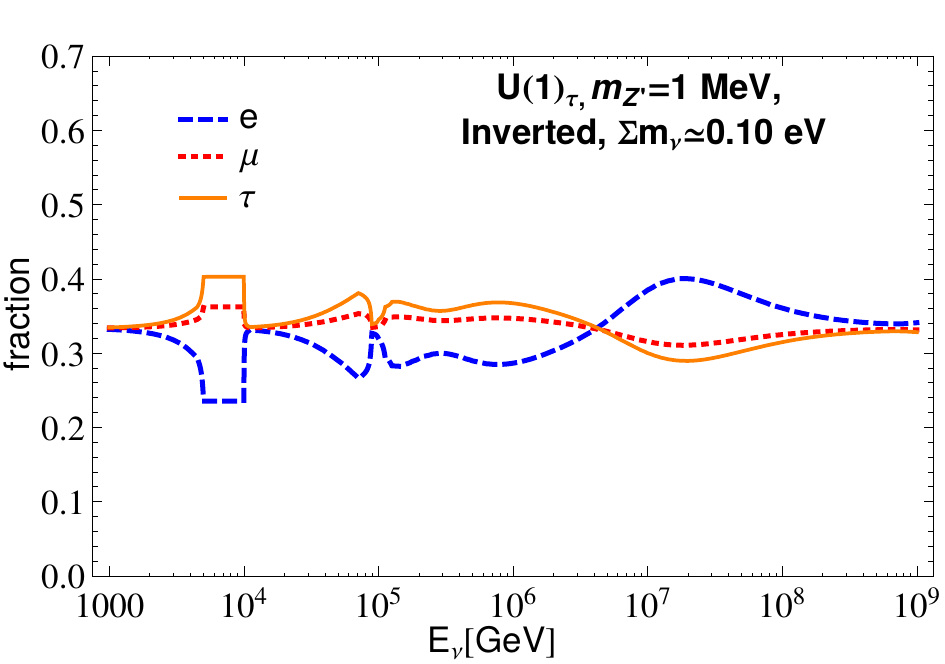}
	\caption{The fractions of neutrinos of each flavor predicted for a $Z'$ associated with the $U(1)_{\mu}$ (upper frame) and $U(1)_{\tau}$ (lower frame) gauge groups. Results are shown for the case of an inverted hierarchy, $\sum m_{\nu}{\approx}0.10$ eV, and $m_{Z'}{=}1$ MeV.}
\label{fig:u1x_frac}
\end{figure}

Thus far, we have restricted our calculations to the case of a $Z'$ associated with the gauge group $U(1)_{\mu-\tau}$. As discussed in Sec.~\ref{models}, however, one could also consider mediators arising from other gauge groups, such as $U(1)_{\mu}$ or $U(1)_{\tau}$. In Fig.~\ref{fig:u1x_spec} we compare the neutrino spectrum predicted for each of these three choices, for the case of the inverted hierarchy with $\sum m_{\nu}{\approx}0.10$ eV and $m_{Z'}{=}1$ MeV. From this comparison, we find that the $U(1)_{\mu-\tau}$ model results in a somewhat smaller degree of absorption at very high energies, despite the fact that more neutrino flavors participate in the scattering. This somewhat counterintuitive result is due to a cancellation between the $\mu$ and $\tau$ contributions to $Q'_{33}$. In Fig~\ref{fig:u1x_frac}, we plot the flavor ratios in each of these scenarios. Here, the magnitude of the impact of the $Z'$ is lessened relative to that predicted in the $U(1)_{\mu-\tau}$ case (see Fig.~\ref{fig:flavor}).


\subsection{Source Distributions}

Up to this point, our calculations have taken all of the neutrino sources to reside at a distance of $z=1$. This distribution was adopted for simplicity, and reflects a plausible average for the distance traveled by a neutrino detected by IceCube. In this subsection, we consider more realistic redshift distributions for the sources of the high-energy neutrinos observed by IceCube. 

\begin{figure}
\includegraphics[width=0.5\textwidth]{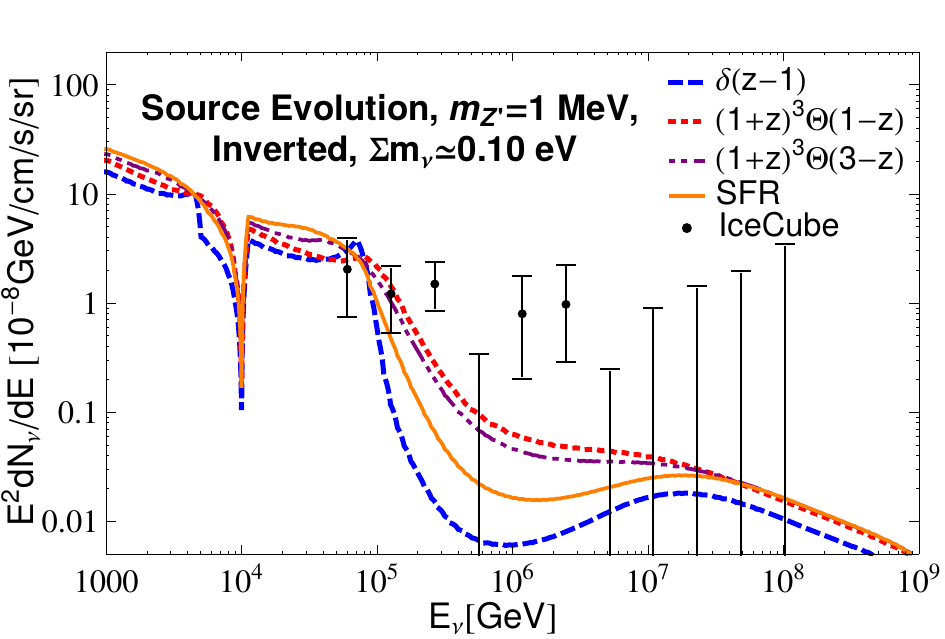}
\caption{A comparison of the neutrino spectra predicted for four source distribution models: all sources at $z=1$ (dashed blue), sources with a constant comoving number density out to $z_{\rm max}=1$ (short dashed red), $z_{\rm max}=3$ (dot-dashed purple), and a distribution of sources that follows the star formation rate (solid orange). Results are shown for the case of an inverted hierarchy, $\sum m_{\nu}{\approx}0.10$ eV, and $m_{Z'}{=}1$ MeV.}
\label{fig:sources}
\end{figure}

\begin{figure}
	\includegraphics[width=0.5\textwidth]{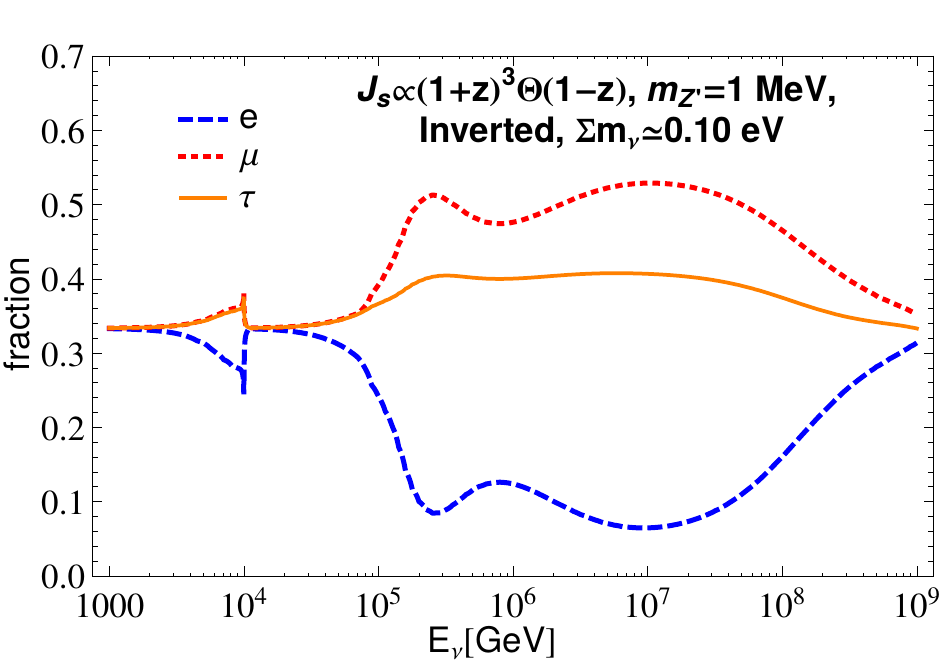}
	\includegraphics[width=0.5\textwidth]{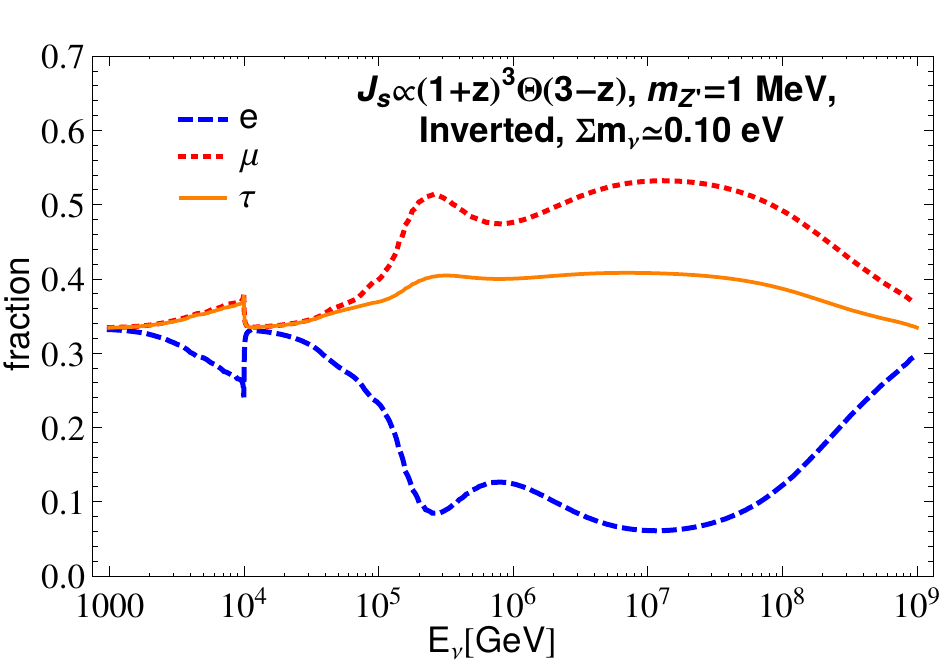}
        \includegraphics[width=0.5\textwidth]{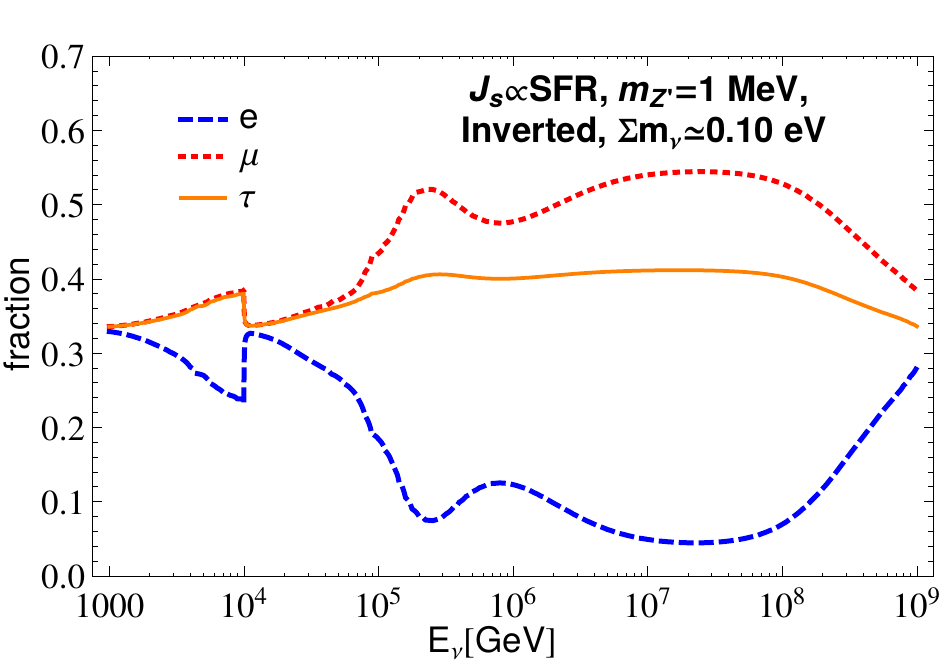}
	\caption{The fraction of neutrinos of each flavor predicted for three source distribution models: sources with a constant comoving number density out to $z_{\rm max}=1$ (upper frame), $z_{\rm max}=3$ (middle frame), and a distribution of sources that follows the star formation rate (lower frame). Results are shown for the case of an inverted hierarchy, $\sum m_{\nu}{\approx}0.10$ eV, and $m_{Z'}{=}1$ MeV.}
	\label{fig:dist}
\end{figure}

The first possibility we consider is a distribution with a constant comoving number density out to $z_{\rm max}$, beyond which no sources exist:
\begin{equation}
J_{\rm comoving}(E_{\nu},z) = A \, E_{\nu}^{\gamma} (1+z)^{3} \,\,\Theta(z_{\rm max}-z),
\end{equation}
where again we take $\gamma=-2.6$. The second distribution that we consider follows the star formation rate~\cite{Yuksel:2008cu}:
\begin{eqnarray}
J_{SFR}(E_{\nu},z)& =  A \, E_{\nu}^{\gamma} (1+z)^{3} \,f_{SFR}(z),
\end{eqnarray}
where
\begin{eqnarray}
f_{SFR}(z)&\equiv \begin{cases} (1+z)^{3.4} & z \leq 1 \\
		2^{3.7} \, (1+z)^{-0.3} & 1 < z < 4 \\
		2^{3.7} \, 5^{3.2}\, (1+z)^{-3.5} & z \geq 4.
\end{cases}
\end{eqnarray}
For sources distributed according to the star formation rate, we consider redshifts up to $z=7$, beyond which contributions to the final spectra are negligible. The neutrino spectra predicted from these distributions are compared in Fig.~\ref{fig:sources}, as are their flavor ratios in Fig.~\ref{fig:dist}. From these figures, we see that our results are qualitatively insensitive to the precise choice of redshift evolution.

\newpage

\section{Summary and Conclusions}
\label{conclusion}

IceCube's recent detection of high-energy astrophysical neutrinos provides us with an opportunity to study the interactions of these particles at higher energies and over longer baselines than are currently possible in laboratory environments. In this paper, we have considered how light ($\sim$1--100 MeV) gauge bosons, with couplings to Standard Model neutrinos, could impact the spectrum and flavors of the neutrinos observed by IceCube. 

New gauge bosons are predicted within a variety of extensions of the Standard Model. Of particular interest is the $Z'$ that arises from the anomaly-free $U(1)_{\mu-\tau}$ gauge group. For masses in the range of $m_{Z'} \sim 1-500$ MeV and a coupling of $g_{Z'} \sim 10^{-3}$, such a particle can explain the measured value of the muon's anomalous magnetic moment, without conflicting with constraints from accelerators or cosmology. For this range of masses and couplings, high-energy astrophysical neutrinos can scatter resonantly with the cosmic neutrino background, leading to absorption features in the spectrum observed at Earth. By measuring the spectrum and the flavor ratios of the extragalactic neutrino flux, IceCube can constrain or provide evidence for such models.

We found the most dramatic effects in models with a very light $Z'$ ($m_{Z'} \lsim 10$ MeV), which induces a significant absorption feature at $E_{\nu} \simeq 5-10\,\,{\rm TeV} \times (m_{Z'}/{\rm MeV})^2$. Although not currently constrained, such a feature could plausibly be measured by IceCube or by next generation neutrino telescopes. Furthermore, in the case of the inverted hierarchy with the lightest neutrino lighter than $\sim$$10^{-3}$ eV, such a $Z'$ can lead to a very broad and deep spectral feature, covering a range of energies between $\sim$\,0.1--10$\,\,{\rm PeV} \times (m_{Z'}/{\rm MeV})^2$. Current IceCube data already excludes this case for a $Z'$ lighter than a few MeV. 

We also emphasize that a $Z'$ can alter the ratios of the neutrino flavors that reach Earth, leading to a different distribution of muon tracks, showers, and tau-unique events at IceCube.  Combining this information with measurements of the neutrino spectrum can significantly extend IceCube's sensitivity to $Z'$ models and to other exotic physics scenarios. 

IceCube's discovery has opened a new window into the interactions of neutrinos at high-energies and over very long baselines. As IceCube and other neutrino telescopes continue to refine their measurements of this population of extragalactic neutrinos, this data will become increasingly sensitive to physics beyond the reach of laboratory experiments. As we have shown, IceCube's current data already excludes a small range of the $Z'$ models considered here. As more data is collected, the range of models within the reach of neutrino telescopes will increase, allowing us to explore a significant fraction of remaining allowed parameter space.

\bigskip \bigskip \bigskip

{\it Acknowledgements}: This work has been supported by the US Department of Energy under Contract No.~DE-FG02-13ER41958. Fermilab is operated by Fermi Research Alliance, LLC, under Contract No.~DE-AC02-07CH11359 with the US Department of Energy. AD is supported by the Fermilab Graduate Student Research Program in Theoretical Physics and in part by NSF Grant No.~PHY-1316792.

\bibliography{icecubezprime}

\begin{thebibliography}{125}%
\makeatletter
\providecommand \@ifxundefined [1]{%
 \@ifx{#1\undefined}
}%
\providecommand \@ifnum [1]{%
 \ifnum #1\expandafter \@firstoftwo
 \else \expandafter \@secondoftwo
 \fi
}%
\providecommand \@ifx [1]{%
 \ifx #1\expandafter \@firstoftwo
 \else \expandafter \@secondoftwo
 \fi
}%
\providecommand \natexlab [1]{#1}%
\providecommand \enquote  [1]{``#1''}%
\providecommand \bibnamefont  [1]{#1}%
\providecommand \bibfnamefont [1]{#1}%
\providecommand \citenamefont [1]{#1}%
\providecommand \href@noop [0]{\@secondoftwo}%
\providecommand \href [0]{\begingroup \@sanitize@url \@href}%
\providecommand \@href[1]{\@@startlink{#1}\@@href}%
\providecommand \@@href[1]{\endgroup#1\@@endlink}%
\providecommand \@sanitize@url [0]{\catcode `\\12\catcode `\$12\catcode
  `\&12\catcode `\#12\catcode `\^12\catcode `\_12\catcode `\%12\relax}%
\providecommand \@@startlink[1]{}%
\providecommand \@@endlink[0]{}%
\providecommand \url  [0]{\begingroup\@sanitize@url \@url }%
\providecommand \@url [1]{\endgroup\@href {#1}{\urlprefix }}%
\providecommand \urlprefix  [0]{URL }%
\providecommand \Eprint [0]{\href }%
\providecommand \doibase [0]{http://dx.doi.org/}%
\providecommand \selectlanguage [0]{\@gobble}%
\providecommand \bibinfo  [0]{\@secondoftwo}%
\providecommand \bibfield  [0]{\@secondoftwo}%
\providecommand \translation [1]{[#1]}%
\providecommand \BibitemOpen [0]{}%
\providecommand \bibitemStop [0]{}%
\providecommand \bibitemNoStop [0]{.\EOS\space}%
\providecommand \EOS [0]{\spacefactor3000\relax}%
\providecommand \BibitemShut  [1]{\csname bibitem#1\endcsname}%
\let\auto@bib@innerbib\@empty
\bibitem [{\citenamefont {Beacom}\ \emph
  {et~al.}(2003{\natexlab{a}})\citenamefont {Beacom}, \citenamefont {Bell},
  \citenamefont {Hooper}, \citenamefont {Pakvasa},\ and\ \citenamefont
  {Weiler}}]{Beacom:2002vi}%
  \BibitemOpen
  \bibfield  {author} {\bibinfo {author} {\bibfnamefont {J.~F.}\ \bibnamefont
  {Beacom}}, \bibinfo {author} {\bibfnamefont {N.~F.}\ \bibnamefont {Bell}},
  \bibinfo {author} {\bibfnamefont {D.}~\bibnamefont {Hooper}}, \bibinfo
  {author} {\bibfnamefont {S.}~\bibnamefont {Pakvasa}}, \ and\ \bibinfo
  {author} {\bibfnamefont {T.~J.}\ \bibnamefont {Weiler}},\ }\href {\doibase
  10.1103/PhysRevLett.90.181301} {\bibfield  {journal} {\bibinfo  {journal}
  {Phys.Rev.Lett.}\ }\textbf {\bibinfo {volume} {90}},\ \bibinfo {pages}
  {181301} (\bibinfo {year} {2003}{\natexlab{a}})},\ \Eprint
  {http://arxiv.org/abs/hep-ph/0211305} {arXiv:hep-ph/0211305 [hep-ph]}
  \BibitemShut {NoStop}%
\bibitem [{\citenamefont {Beacom}\ \emph
  {et~al.}(2003{\natexlab{b}})\citenamefont {Beacom}, \citenamefont {Bell},
  \citenamefont {Hooper}, \citenamefont {Pakvasa},\ and\ \citenamefont
  {Weiler}}]{Beacom:2003nh}%
  \BibitemOpen
  \bibfield  {author} {\bibinfo {author} {\bibfnamefont {J.~F.}\ \bibnamefont
  {Beacom}}, \bibinfo {author} {\bibfnamefont {N.~F.}\ \bibnamefont {Bell}},
  \bibinfo {author} {\bibfnamefont {D.}~\bibnamefont {Hooper}}, \bibinfo
  {author} {\bibfnamefont {S.}~\bibnamefont {Pakvasa}}, \ and\ \bibinfo
  {author} {\bibfnamefont {T.~J.}\ \bibnamefont {Weiler}},\ }\href {\doibase
  10.1103/PhysRevD.68.093005, 10.1103/PhysRevD.72.019901} {\bibfield  {journal}
  {\bibinfo  {journal} {Phys.Rev.}\ }\textbf {\bibinfo {volume} {D68}},\
  \bibinfo {pages} {093005} (\bibinfo {year} {2003}{\natexlab{b}})},\ \Eprint
  {http://arxiv.org/abs/hep-ph/0307025} {arXiv:hep-ph/0307025 [hep-ph]}
  \BibitemShut {NoStop}%
\bibitem [{\citenamefont {Beacom}\ \emph
  {et~al.}(2004{\natexlab{a}})\citenamefont {Beacom}, \citenamefont {Bell},
  \citenamefont {Hooper}, \citenamefont {Learned}, \citenamefont {Pakvasa}
  \emph {et~al.}}]{Beacom:2003eu}%
  \BibitemOpen
  \bibfield  {author} {\bibinfo {author} {\bibfnamefont {J.~F.}\ \bibnamefont
  {Beacom}}, \bibinfo {author} {\bibfnamefont {N.~F.}\ \bibnamefont {Bell}},
  \bibinfo {author} {\bibfnamefont {D.}~\bibnamefont {Hooper}}, \bibinfo
  {author} {\bibfnamefont {J.~G.}\ \bibnamefont {Learned}}, \bibinfo {author}
  {\bibfnamefont {S.}~\bibnamefont {Pakvasa}},  \emph {et~al.},\ }\href
  {\doibase 10.1103/PhysRevLett.92.011101} {\bibfield  {journal} {\bibinfo
  {journal} {Phys.Rev.Lett.}\ }\textbf {\bibinfo {volume} {92}},\ \bibinfo
  {pages} {011101} (\bibinfo {year} {2004}{\natexlab{a}})},\ \Eprint
  {http://arxiv.org/abs/hep-ph/0307151} {arXiv:hep-ph/0307151 [hep-ph]}
  \BibitemShut {NoStop}%
\bibitem [{\citenamefont {Beacom}\ \emph
  {et~al.}(2004{\natexlab{b}})\citenamefont {Beacom}, \citenamefont {Bell},
  \citenamefont {Hooper}, \citenamefont {Pakvasa},\ and\ \citenamefont
  {Weiler}}]{Beacom:2003zg}%
  \BibitemOpen
  \bibfield  {author} {\bibinfo {author} {\bibfnamefont {J.~F.}\ \bibnamefont
  {Beacom}}, \bibinfo {author} {\bibfnamefont {N.~F.}\ \bibnamefont {Bell}},
  \bibinfo {author} {\bibfnamefont {D.}~\bibnamefont {Hooper}}, \bibinfo
  {author} {\bibfnamefont {S.}~\bibnamefont {Pakvasa}}, \ and\ \bibinfo
  {author} {\bibfnamefont {T.~J.}\ \bibnamefont {Weiler}},\ }\href {\doibase
  10.1103/PhysRevD.69.017303} {\bibfield  {journal} {\bibinfo  {journal}
  {Phys.Rev.}\ }\textbf {\bibinfo {volume} {D69}},\ \bibinfo {pages} {017303}
  (\bibinfo {year} {2004}{\natexlab{b}})},\ \Eprint
  {http://arxiv.org/abs/hep-ph/0309267} {arXiv:hep-ph/0309267 [hep-ph]}
  \BibitemShut {NoStop}%
\bibitem [{\citenamefont {Pagliaroli}\ \emph {et~al.}(2015)\citenamefont
  {Pagliaroli}, \citenamefont {Palladino}, \citenamefont {Vissani},\ and\
  \citenamefont {Villante}}]{Pagliaroli:2015rca}%
  \BibitemOpen
  \bibfield  {author} {\bibinfo {author} {\bibfnamefont {G.}~\bibnamefont
  {Pagliaroli}}, \bibinfo {author} {\bibfnamefont {A.}~\bibnamefont
  {Palladino}}, \bibinfo {author} {\bibfnamefont {F.}~\bibnamefont {Vissani}},
  \ and\ \bibinfo {author} {\bibfnamefont {F.}~\bibnamefont {Villante}},\
  }\href@noop {} {\  (\bibinfo {year} {2015})},\ \Eprint
  {http://arxiv.org/abs/1506.02624} {arXiv:1506.02624 [hep-ph]} \BibitemShut
  {NoStop}%
\bibitem [{\citenamefont {Bustamante}\ \emph {et~al.}(2015)\citenamefont
  {Bustamante}, \citenamefont {Beacom},\ and\ \citenamefont
  {Winter}}]{Bustamante:2015waa}%
  \BibitemOpen
  \bibfield  {author} {\bibinfo {author} {\bibfnamefont {M.}~\bibnamefont
  {Bustamante}}, \bibinfo {author} {\bibfnamefont {J.~F.}\ \bibnamefont
  {Beacom}}, \ and\ \bibinfo {author} {\bibfnamefont {W.}~\bibnamefont
  {Winter}},\ }\href@noop {} {\  (\bibinfo {year} {2015})},\ \Eprint
  {http://arxiv.org/abs/1506.02645} {arXiv:1506.02645 [astro-ph.HE]}
  \BibitemShut {NoStop}%
\bibitem [{\citenamefont {Aeikens}\ \emph {et~al.}(2014)\citenamefont
  {Aeikens}, \citenamefont {P{\"a}s}, \citenamefont {Pakvasa},\ and\
  \citenamefont {Sicking}}]{Aeikens:2014yga}%
  \BibitemOpen
  \bibfield  {author} {\bibinfo {author} {\bibfnamefont {E.}~\bibnamefont
  {Aeikens}}, \bibinfo {author} {\bibfnamefont {H.}~\bibnamefont {P{\"a}s}},
  \bibinfo {author} {\bibfnamefont {S.}~\bibnamefont {Pakvasa}}, \ and\
  \bibinfo {author} {\bibfnamefont {P.}~\bibnamefont {Sicking}},\ }\href@noop
  {} {\  (\bibinfo {year} {2014})},\ \Eprint {http://arxiv.org/abs/1410.0408}
  {arXiv:1410.0408 [hep-ph]} \BibitemShut {NoStop}%
\bibitem [{\citenamefont {Hollander}(2013)}]{Hollander:2013im}%
  \BibitemOpen
  \bibfield  {author} {\bibinfo {author} {\bibfnamefont {D.}~\bibnamefont
  {Hollander}},\ }\href@noop {} {\  (\bibinfo {year} {2013})},\ \Eprint
  {http://arxiv.org/abs/1301.5313} {arXiv:1301.5313 [hep-ph]} \BibitemShut
  {NoStop}%
\bibitem [{\citenamefont {Fu}\ \emph {et~al.}(2012)\citenamefont {Fu},
  \citenamefont {Ho},\ and\ \citenamefont {Weiler}}]{Fu:2012zr}%
  \BibitemOpen
  \bibfield  {author} {\bibinfo {author} {\bibfnamefont {L.}~\bibnamefont
  {Fu}}, \bibinfo {author} {\bibfnamefont {C.~M.}\ \bibnamefont {Ho}}, \ and\
  \bibinfo {author} {\bibfnamefont {T.~J.}\ \bibnamefont {Weiler}},\ }\href
  {\doibase 10.1016/j.physletb.2012.11.011} {\bibfield  {journal} {\bibinfo
  {journal} {Phys.Lett.}\ }\textbf {\bibinfo {volume} {B718}},\ \bibinfo
  {pages} {558} (\bibinfo {year} {2012})},\ \Eprint
  {http://arxiv.org/abs/1209.5382} {arXiv:1209.5382 [hep-ph]} \BibitemShut
  {NoStop}%
\bibitem [{\citenamefont {Bhattacharya}\ \emph
  {et~al.}(2010{\natexlab{a}})\citenamefont {Bhattacharya}, \citenamefont
  {Choubey}, \citenamefont {Gandhi},\ and\ \citenamefont
  {Watanabe}}]{Bhattacharya:2010xj}%
  \BibitemOpen
  \bibfield  {author} {\bibinfo {author} {\bibfnamefont {A.}~\bibnamefont
  {Bhattacharya}}, \bibinfo {author} {\bibfnamefont {S.}~\bibnamefont
  {Choubey}}, \bibinfo {author} {\bibfnamefont {R.}~\bibnamefont {Gandhi}}, \
  and\ \bibinfo {author} {\bibfnamefont {A.}~\bibnamefont {Watanabe}},\ }\href
  {\doibase 10.1088/1475-7516/2010/09/009} {\bibfield  {journal} {\bibinfo
  {journal} {JCAP}\ }\textbf {\bibinfo {volume} {1009}},\ \bibinfo {pages}
  {009} (\bibinfo {year} {2010}{\natexlab{a}})},\ \Eprint
  {http://arxiv.org/abs/1006.3082} {arXiv:1006.3082 [hep-ph]} \BibitemShut
  {NoStop}%
\bibitem [{\citenamefont {Bustamante}\ \emph {et~al.}(2010)\citenamefont
  {Bustamante}, \citenamefont {Gago},\ and\ \citenamefont
  {Pena-Garay}}]{Bustamante:2010nq}%
  \BibitemOpen
  \bibfield  {author} {\bibinfo {author} {\bibfnamefont {M.}~\bibnamefont
  {Bustamante}}, \bibinfo {author} {\bibfnamefont {A.}~\bibnamefont {Gago}}, \
  and\ \bibinfo {author} {\bibfnamefont {C.}~\bibnamefont {Pena-Garay}},\
  }\href {\doibase 10.1007/JHEP04(2010)066} {\bibfield  {journal} {\bibinfo
  {journal} {JHEP}\ }\textbf {\bibinfo {volume} {1004}},\ \bibinfo {pages}
  {066} (\bibinfo {year} {2010})},\ \Eprint {http://arxiv.org/abs/1001.4878}
  {arXiv:1001.4878 [hep-ph]} \BibitemShut {NoStop}%
\bibitem [{\citenamefont {Bhattacharya}\ \emph
  {et~al.}(2010{\natexlab{b}})\citenamefont {Bhattacharya}, \citenamefont
  {Choubey}, \citenamefont {Gandhi},\ and\ \citenamefont
  {Watanabe}}]{Bhattacharya:2009tx}%
  \BibitemOpen
  \bibfield  {author} {\bibinfo {author} {\bibfnamefont {A.}~\bibnamefont
  {Bhattacharya}}, \bibinfo {author} {\bibfnamefont {S.}~\bibnamefont
  {Choubey}}, \bibinfo {author} {\bibfnamefont {R.}~\bibnamefont {Gandhi}}, \
  and\ \bibinfo {author} {\bibfnamefont {A.}~\bibnamefont {Watanabe}},\ }\href
  {\doibase 10.1016/j.physletb.2010.04.078} {\bibfield  {journal} {\bibinfo
  {journal} {Phys.Lett.}\ }\textbf {\bibinfo {volume} {B690}},\ \bibinfo
  {pages} {42} (\bibinfo {year} {2010}{\natexlab{b}})},\ \Eprint
  {http://arxiv.org/abs/0910.4396} {arXiv:0910.4396 [hep-ph]} \BibitemShut
  {NoStop}%
\bibitem [{\citenamefont {Xing}\ and\ \citenamefont
  {Zhou}(2008)}]{Xing:2008fg}%
  \BibitemOpen
  \bibfield  {author} {\bibinfo {author} {\bibfnamefont {Z.-z.}\ \bibnamefont
  {Xing}}\ and\ \bibinfo {author} {\bibfnamefont {S.}~\bibnamefont {Zhou}},\
  }\href {\doibase 10.1016/j.physletb.2008.07.011} {\bibfield  {journal}
  {\bibinfo  {journal} {Phys.Lett.}\ }\textbf {\bibinfo {volume} {B666}},\
  \bibinfo {pages} {166} (\bibinfo {year} {2008})},\ \Eprint
  {http://arxiv.org/abs/0804.3512} {arXiv:0804.3512 [hep-ph]} \BibitemShut
  {NoStop}%
\bibitem [{\citenamefont {Hooper}\ \emph
  {et~al.}(2005{\natexlab{a}})\citenamefont {Hooper}, \citenamefont {Morgan},\
  and\ \citenamefont {Winstanley}}]{Hooper:2005jp}%
  \BibitemOpen
  \bibfield  {author} {\bibinfo {author} {\bibfnamefont {D.}~\bibnamefont
  {Hooper}}, \bibinfo {author} {\bibfnamefont {D.}~\bibnamefont {Morgan}}, \
  and\ \bibinfo {author} {\bibfnamefont {E.}~\bibnamefont {Winstanley}},\
  }\href {\doibase 10.1103/PhysRevD.72.065009} {\bibfield  {journal} {\bibinfo
  {journal} {Phys.Rev.}\ }\textbf {\bibinfo {volume} {D72}},\ \bibinfo {pages}
  {065009} (\bibinfo {year} {2005}{\natexlab{a}})},\ \Eprint
  {http://arxiv.org/abs/hep-ph/0506091} {arXiv:hep-ph/0506091 [hep-ph]}
  \BibitemShut {NoStop}%
\bibitem [{\citenamefont {Hooper}\ \emph
  {et~al.}(2005{\natexlab{b}})\citenamefont {Hooper}, \citenamefont {Morgan},\
  and\ \citenamefont {Winstanley}}]{Hooper:2004xr}%
  \BibitemOpen
  \bibfield  {author} {\bibinfo {author} {\bibfnamefont {D.}~\bibnamefont
  {Hooper}}, \bibinfo {author} {\bibfnamefont {D.}~\bibnamefont {Morgan}}, \
  and\ \bibinfo {author} {\bibfnamefont {E.}~\bibnamefont {Winstanley}},\
  }\href {\doibase 10.1016/j.physletb.2005.01.034} {\bibfield  {journal}
  {\bibinfo  {journal} {Phys.Lett.}\ }\textbf {\bibinfo {volume} {B609}},\
  \bibinfo {pages} {206} (\bibinfo {year} {2005}{\natexlab{b}})},\ \Eprint
  {http://arxiv.org/abs/hep-ph/0410094} {arXiv:hep-ph/0410094 [hep-ph]}
  \BibitemShut {NoStop}%
\bibitem [{\citenamefont {Argüelles}\ \emph {et~al.}(2015)\citenamefont
  {Argüelles}, \citenamefont {Katori},\ and\ \citenamefont
  {Salvado}}]{Arguelles:2015dca}%
  \BibitemOpen
  \bibfield  {author} {\bibinfo {author} {\bibfnamefont {C.~A.}\ \bibnamefont
  {Argüelles}}, \bibinfo {author} {\bibfnamefont {T.}~\bibnamefont {Katori}},
  \ and\ \bibinfo {author} {\bibfnamefont {J.}~\bibnamefont {Salvado}},\
  }\href@noop {} {\  (\bibinfo {year} {2015})},\ \Eprint
  {http://arxiv.org/abs/1506.02043} {arXiv:1506.02043 [hep-ph]} \BibitemShut
  {NoStop}%
\bibitem [{\citenamefont {Farzan}\ and\ \citenamefont
  {Palomares-Ruiz}(2014)}]{Farzan:2014gza}%
  \BibitemOpen
  \bibfield  {author} {\bibinfo {author} {\bibfnamefont {Y.}~\bibnamefont
  {Farzan}}\ and\ \bibinfo {author} {\bibfnamefont {S.}~\bibnamefont
  {Palomares-Ruiz}},\ }\href {\doibase 10.1088/1475-7516/2014/06/014}
  {\bibfield  {journal} {\bibinfo  {journal} {JCAP}\ }\textbf {\bibinfo
  {volume} {1406}},\ \bibinfo {pages} {014} (\bibinfo {year} {2014})},\ \Eprint
  {http://arxiv.org/abs/1401.7019} {arXiv:1401.7019 [hep-ph]} \BibitemShut
  {NoStop}%
\bibitem [{\citenamefont {Chatterjee}\ \emph {et~al.}(2014)\citenamefont
  {Chatterjee}, \citenamefont {Devi}, \citenamefont {Ghosh}, \citenamefont
  {Moharana},\ and\ \citenamefont {Raut}}]{Chatterjee:2013tza}%
  \BibitemOpen
  \bibfield  {author} {\bibinfo {author} {\bibfnamefont {A.}~\bibnamefont
  {Chatterjee}}, \bibinfo {author} {\bibfnamefont {M.~M.}\ \bibnamefont
  {Devi}}, \bibinfo {author} {\bibfnamefont {M.}~\bibnamefont {Ghosh}},
  \bibinfo {author} {\bibfnamefont {R.}~\bibnamefont {Moharana}}, \ and\
  \bibinfo {author} {\bibfnamefont {S.~K.}\ \bibnamefont {Raut}},\ }\href
  {\doibase 10.1103/PhysRevD.90.073003} {\bibfield  {journal} {\bibinfo
  {journal} {Phys. Rev.}\ }\textbf {\bibinfo {volume} {D90}},\ \bibinfo {pages}
  {073003} (\bibinfo {year} {2014})},\ \Eprint {http://arxiv.org/abs/1312.6593}
  {arXiv:1312.6593 [hep-ph]} \BibitemShut {NoStop}%
\bibitem [{\citenamefont {Illana}\ \emph {et~al.}(2015)\citenamefont {Illana},
  \citenamefont {Masip},\ and\ \citenamefont {Meloni}}]{Illana:2014bda}%
  \BibitemOpen
  \bibfield  {author} {\bibinfo {author} {\bibfnamefont {J.~I.}\ \bibnamefont
  {Illana}}, \bibinfo {author} {\bibfnamefont {M.}~\bibnamefont {Masip}}, \
  and\ \bibinfo {author} {\bibfnamefont {D.}~\bibnamefont {Meloni}},\ }\href
  {\doibase 10.1016/j.astropartphys.2014.12.004} {\bibfield  {journal}
  {\bibinfo  {journal} {Astropart.Phys.}\ }\textbf {\bibinfo {volume} {65}},\
  \bibinfo {pages} {64} (\bibinfo {year} {2015})},\ \Eprint
  {http://arxiv.org/abs/1410.3208} {arXiv:1410.3208 [hep-ph]} \BibitemShut
  {NoStop}%
\bibitem [{\citenamefont {Alvarez-Muniz}\ \emph
  {et~al.}(2002{\natexlab{a}})\citenamefont {Alvarez-Muniz}, \citenamefont
  {Feng}, \citenamefont {Halzen}, \citenamefont {Han},\ and\ \citenamefont
  {Hooper}}]{AlvarezMuniz:2002ga}%
  \BibitemOpen
  \bibfield  {author} {\bibinfo {author} {\bibfnamefont {J.}~\bibnamefont
  {Alvarez-Muniz}}, \bibinfo {author} {\bibfnamefont {J.~L.}\ \bibnamefont
  {Feng}}, \bibinfo {author} {\bibfnamefont {F.}~\bibnamefont {Halzen}},
  \bibinfo {author} {\bibfnamefont {T.}~\bibnamefont {Han}}, \ and\ \bibinfo
  {author} {\bibfnamefont {D.}~\bibnamefont {Hooper}},\ }\href {\doibase
  10.1103/PhysRevD.65.124015} {\bibfield  {journal} {\bibinfo  {journal}
  {Phys.Rev.}\ }\textbf {\bibinfo {volume} {D65}},\ \bibinfo {pages} {124015}
  (\bibinfo {year} {2002}{\natexlab{a}})},\ \Eprint
  {http://arxiv.org/abs/hep-ph/0202081} {arXiv:hep-ph/0202081 [hep-ph]}
  \BibitemShut {NoStop}%
\bibitem [{\citenamefont {Alvarez-Muniz}\ \emph
  {et~al.}(2002{\natexlab{b}})\citenamefont {Alvarez-Muniz}, \citenamefont
  {Halzen}, \citenamefont {Han},\ and\ \citenamefont
  {Hooper}}]{AlvarezMuniz:2001mk}%
  \BibitemOpen
  \bibfield  {author} {\bibinfo {author} {\bibfnamefont {J.}~\bibnamefont
  {Alvarez-Muniz}}, \bibinfo {author} {\bibfnamefont {F.}~\bibnamefont
  {Halzen}}, \bibinfo {author} {\bibfnamefont {T.}~\bibnamefont {Han}}, \ and\
  \bibinfo {author} {\bibfnamefont {D.}~\bibnamefont {Hooper}},\ }\href
  {\doibase 10.1103/PhysRevLett.88.021301} {\bibfield  {journal} {\bibinfo
  {journal} {Phys.Rev.Lett.}\ }\textbf {\bibinfo {volume} {88}},\ \bibinfo
  {pages} {021301} (\bibinfo {year} {2002}{\natexlab{b}})},\ \Eprint
  {http://arxiv.org/abs/hep-ph/0107057} {arXiv:hep-ph/0107057 [hep-ph]}
  \BibitemShut {NoStop}%
\bibitem [{\citenamefont {Friess}\ \emph {et~al.}(2002)\citenamefont {Friess},
  \citenamefont {Han},\ and\ \citenamefont {Hooper}}]{Friess:2002cc}%
  \BibitemOpen
  \bibfield  {author} {\bibinfo {author} {\bibfnamefont {J.~J.}\ \bibnamefont
  {Friess}}, \bibinfo {author} {\bibfnamefont {T.}~\bibnamefont {Han}}, \ and\
  \bibinfo {author} {\bibfnamefont {D.}~\bibnamefont {Hooper}},\ }\href
  {\doibase 10.1016/S0370-2693(02)02728-4} {\bibfield  {journal} {\bibinfo
  {journal} {Phys.Lett.}\ }\textbf {\bibinfo {volume} {B547}},\ \bibinfo
  {pages} {31} (\bibinfo {year} {2002})},\ \Eprint
  {http://arxiv.org/abs/hep-ph/0204112} {arXiv:hep-ph/0204112 [hep-ph]}
  \BibitemShut {NoStop}%
\bibitem [{\citenamefont {Han}\ and\ \citenamefont
  {Hooper}(2004)}]{Han:2004kq}%
  \BibitemOpen
  \bibfield  {author} {\bibinfo {author} {\bibfnamefont {T.}~\bibnamefont
  {Han}}\ and\ \bibinfo {author} {\bibfnamefont {D.}~\bibnamefont {Hooper}},\
  }\href {\doibase 10.1088/1367-2630/6/1/150} {\bibfield  {journal} {\bibinfo
  {journal} {New J.Phys.}\ }\textbf {\bibinfo {volume} {6}},\ \bibinfo {pages}
  {150} (\bibinfo {year} {2004})},\ \Eprint
  {http://arxiv.org/abs/hep-ph/0408348} {arXiv:hep-ph/0408348 [hep-ph]}
  \BibitemShut {NoStop}%
\bibitem [{\citenamefont {Dutta}\ \emph {et~al.}(2015)\citenamefont {Dutta},
  \citenamefont {Gao}, \citenamefont {Li}, \citenamefont {Rott},\ and\
  \citenamefont {Strigari}}]{Dutta:2015dka}%
  \BibitemOpen
  \bibfield  {author} {\bibinfo {author} {\bibfnamefont {B.}~\bibnamefont
  {Dutta}}, \bibinfo {author} {\bibfnamefont {Y.}~\bibnamefont {Gao}}, \bibinfo
  {author} {\bibfnamefont {T.}~\bibnamefont {Li}}, \bibinfo {author}
  {\bibfnamefont {C.}~\bibnamefont {Rott}}, \ and\ \bibinfo {author}
  {\bibfnamefont {L.~E.}\ \bibnamefont {Strigari}},\ }\href@noop {} {\
  (\bibinfo {year} {2015})},\ \Eprint {http://arxiv.org/abs/1505.00028}
  {arXiv:1505.00028 [hep-ph]} \BibitemShut {NoStop}%
\bibitem [{\citenamefont {Dey}\ and\ \citenamefont
  {Mohanty}(2015)}]{Dey:2015eaa}%
  \BibitemOpen
  \bibfield  {author} {\bibinfo {author} {\bibfnamefont {U.~K.}\ \bibnamefont
  {Dey}}\ and\ \bibinfo {author} {\bibfnamefont {S.}~\bibnamefont {Mohanty}},\
  }\href@noop {} {\  (\bibinfo {year} {2015})},\ \Eprint
  {http://arxiv.org/abs/1505.01037} {arXiv:1505.01037 [hep-ph]} \BibitemShut
  {NoStop}%
\bibitem [{\citenamefont {Hooper}(2007)}]{Hooper:2007jr}%
  \BibitemOpen
  \bibfield  {author} {\bibinfo {author} {\bibfnamefont {D.}~\bibnamefont
  {Hooper}},\ }\href {\doibase 10.1103/PhysRevD.75.123001} {\bibfield
  {journal} {\bibinfo  {journal} {Phys.Rev.}\ }\textbf {\bibinfo {volume}
  {D75}},\ \bibinfo {pages} {123001} (\bibinfo {year} {2007})},\ \Eprint
  {http://arxiv.org/abs/hep-ph/0701194} {arXiv:hep-ph/0701194 [hep-ph]}
  \BibitemShut {NoStop}%
\bibitem [{\citenamefont {Araki}\ \emph
  {et~al.}(2015{\natexlab{a}})\citenamefont {Araki}, \citenamefont {Kaneko},
  \citenamefont {Konishi}, \citenamefont {Ota}, \citenamefont {Sato} \emph
  {et~al.}}]{Araki:2014ona}%
  \BibitemOpen
  \bibfield  {author} {\bibinfo {author} {\bibfnamefont {T.}~\bibnamefont
  {Araki}}, \bibinfo {author} {\bibfnamefont {F.}~\bibnamefont {Kaneko}},
  \bibinfo {author} {\bibfnamefont {Y.}~\bibnamefont {Konishi}}, \bibinfo
  {author} {\bibfnamefont {T.}~\bibnamefont {Ota}}, \bibinfo {author}
  {\bibfnamefont {J.}~\bibnamefont {Sato}},  \emph {et~al.},\ }\href {\doibase
  10.1103/PhysRevD.91.037301} {\bibfield  {journal} {\bibinfo  {journal}
  {Phys.Rev.}\ }\textbf {\bibinfo {volume} {D91}},\ \bibinfo {pages} {037301}
  (\bibinfo {year} {2015}{\natexlab{a}})},\ \Eprint
  {http://arxiv.org/abs/1409.4180} {arXiv:1409.4180 [hep-ph]} \BibitemShut
  {NoStop}%
\bibitem [{\citenamefont {Ng}\ and\ \citenamefont {Beacom}(2014)}]{Ng:2014pca}%
  \BibitemOpen
  \bibfield  {author} {\bibinfo {author} {\bibfnamefont {K.~C.~Y.}\
  \bibnamefont {Ng}}\ and\ \bibinfo {author} {\bibfnamefont {J.~F.}\
  \bibnamefont {Beacom}},\ }\href {\doibase 10.1103/PhysRevD.90.065035,
  10.1103/PhysRevD.90.089904} {\bibfield  {journal} {\bibinfo  {journal}
  {Phys.Rev.}\ }\textbf {\bibinfo {volume} {D90}},\ \bibinfo {pages} {065035}
  (\bibinfo {year} {2014})},\ \Eprint {http://arxiv.org/abs/1404.2288}
  {arXiv:1404.2288 [astro-ph.HE]} \BibitemShut {NoStop}%
\bibitem [{\citenamefont {Araki}\ \emph
  {et~al.}(2015{\natexlab{b}})\citenamefont {Araki}, \citenamefont {Kaneko},
  \citenamefont {Konishi}, \citenamefont {Ota}, \citenamefont {Sato} \emph
  {et~al.}}]{Araki:2015dia}%
  \BibitemOpen
  \bibfield  {author} {\bibinfo {author} {\bibfnamefont {T.}~\bibnamefont
  {Araki}}, \bibinfo {author} {\bibfnamefont {F.}~\bibnamefont {Kaneko}},
  \bibinfo {author} {\bibfnamefont {Y.}~\bibnamefont {Konishi}}, \bibinfo
  {author} {\bibfnamefont {T.}~\bibnamefont {Ota}}, \bibinfo {author}
  {\bibfnamefont {J.}~\bibnamefont {Sato}},  \emph {et~al.},\ }\href@noop {} {\
   (\bibinfo {year} {2015}{\natexlab{b}})},\ \Eprint
  {http://arxiv.org/abs/1505.01284} {arXiv:1505.01284 [hep-ph]} \BibitemShut
  {NoStop}%
\bibitem [{\citenamefont {Kamada}\ and\ \citenamefont
  {Yu}(2015)}]{Kamada:2015era}%
  \BibitemOpen
  \bibfield  {author} {\bibinfo {author} {\bibfnamefont {A.}~\bibnamefont
  {Kamada}}\ and\ \bibinfo {author} {\bibfnamefont {H.-B.}\ \bibnamefont
  {Yu}},\ }\href@noop {} {\  (\bibinfo {year} {2015})},\ \Eprint
  {http://arxiv.org/abs/1504.00711} {arXiv:1504.00711 [hep-ph]} \BibitemShut
  {NoStop}%
\bibitem [{\citenamefont {Ibe}\ and\ \citenamefont
  {Kaneta}(2014)}]{Ibe:2014pja}%
  \BibitemOpen
  \bibfield  {author} {\bibinfo {author} {\bibfnamefont {M.}~\bibnamefont
  {Ibe}}\ and\ \bibinfo {author} {\bibfnamefont {K.}~\bibnamefont {Kaneta}},\
  }\href {\doibase 10.1103/PhysRevD.90.053011} {\bibfield  {journal} {\bibinfo
  {journal} {Phys.Rev.}\ }\textbf {\bibinfo {volume} {D90}},\ \bibinfo {pages}
  {053011} (\bibinfo {year} {2014})},\ \Eprint {http://arxiv.org/abs/1407.2848}
  {arXiv:1407.2848 [hep-ph]} \BibitemShut {NoStop}%
\bibitem [{\citenamefont {Ioka}\ and\ \citenamefont
  {Murase}(2014)}]{Ioka:2014kca}%
  \BibitemOpen
  \bibfield  {author} {\bibinfo {author} {\bibfnamefont {K.}~\bibnamefont
  {Ioka}}\ and\ \bibinfo {author} {\bibfnamefont {K.}~\bibnamefont {Murase}},\
  }\href {\doibase 10.1093/ptep/ptu090} {\bibfield  {journal} {\bibinfo
  {journal} {PTEP}\ }\textbf {\bibinfo {volume} {2014}},\ \bibinfo {pages}
  {061E01} (\bibinfo {year} {2014})},\ \Eprint {http://arxiv.org/abs/1404.2279}
  {arXiv:1404.2279 [astro-ph.HE]} \BibitemShut {NoStop}%
\bibitem [{\citenamefont {Farzan}(2015)}]{Farzan:2015doa}%
  \BibitemOpen
  \bibfield  {author} {\bibinfo {author} {\bibfnamefont {Y.}~\bibnamefont
  {Farzan}},\ }\href {\doibase 10.1016/j.physletb.2015.07.015} {\  (\bibinfo
  {year} {2015}),\ 10.1016/j.physletb.2015.07.015},\ \Eprint
  {http://arxiv.org/abs/1505.06906} {arXiv:1505.06906 [hep-ph]} \BibitemShut
  {NoStop}%
\bibitem [{\citenamefont {Cherry}\ \emph {et~al.}(2014)\citenamefont {Cherry},
  \citenamefont {Friedland},\ and\ \citenamefont {Shoemaker}}]{Cherry:2014xra}%
  \BibitemOpen
  \bibfield  {author} {\bibinfo {author} {\bibfnamefont {J.~F.}\ \bibnamefont
  {Cherry}}, \bibinfo {author} {\bibfnamefont {A.}~\bibnamefont {Friedland}}, \
  and\ \bibinfo {author} {\bibfnamefont {I.~M.}\ \bibnamefont {Shoemaker}},\
  }\href@noop {} {\  (\bibinfo {year} {2014})},\ \Eprint
  {http://arxiv.org/abs/1411.1071} {arXiv:1411.1071 [hep-ph]} \BibitemShut
  {NoStop}%
\bibitem [{\citenamefont {Aartsen}\ \emph
  {et~al.}(2013{\natexlab{a}})\citenamefont {Aartsen} \emph
  {et~al.}}]{Aartsen:2013bka}%
  \BibitemOpen
  \bibfield  {author} {\bibinfo {author} {\bibfnamefont {M.}~\bibnamefont
  {Aartsen}} \emph {et~al.} (\bibinfo {collaboration} {IceCube}),\ }\href
  {\doibase 10.1103/PhysRevLett.111.021103} {\bibfield  {journal} {\bibinfo
  {journal} {Phys.Rev.Lett.}\ }\textbf {\bibinfo {volume} {111}},\ \bibinfo
  {pages} {021103} (\bibinfo {year} {2013}{\natexlab{a}})},\ \Eprint
  {http://arxiv.org/abs/1304.5356} {arXiv:1304.5356 [astro-ph.HE]} \BibitemShut
  {NoStop}%
\bibitem [{\citenamefont {Aartsen}\ \emph
  {et~al.}(2013{\natexlab{b}})\citenamefont {Aartsen} \emph
  {et~al.}}]{Aartsen:2013jdh}%
  \BibitemOpen
  \bibfield  {author} {\bibinfo {author} {\bibfnamefont {M.}~\bibnamefont
  {Aartsen}} \emph {et~al.} (\bibinfo {collaboration} {IceCube}),\ }\href
  {\doibase 10.1126/science.1242856} {\bibfield  {journal} {\bibinfo  {journal}
  {Science}\ }\textbf {\bibinfo {volume} {342}},\ \bibinfo {pages} {1242856}
  (\bibinfo {year} {2013}{\natexlab{b}})},\ \Eprint
  {http://arxiv.org/abs/1311.5238} {arXiv:1311.5238 [astro-ph.HE]} \BibitemShut
  {NoStop}%
\bibitem [{\citenamefont {Aartsen}\ \emph {et~al.}(2014)\citenamefont {Aartsen}
  \emph {et~al.}}]{Aartsen:2014gkd}%
  \BibitemOpen
  \bibfield  {author} {\bibinfo {author} {\bibfnamefont {M.}~\bibnamefont
  {Aartsen}} \emph {et~al.} (\bibinfo {collaboration} {IceCube}),\ }\href
  {\doibase 10.1103/PhysRevLett.113.101101} {\bibfield  {journal} {\bibinfo
  {journal} {Phys.Rev.Lett.}\ }\textbf {\bibinfo {volume} {113}},\ \bibinfo
  {pages} {101101} (\bibinfo {year} {2014})},\ \Eprint
  {http://arxiv.org/abs/1405.5303} {arXiv:1405.5303 [astro-ph.HE]} \BibitemShut
  {NoStop}%
\bibitem [{\citenamefont {Ahlers}\ \emph {et~al.}(2015)\citenamefont {Ahlers},
  \citenamefont {Bai}, \citenamefont {Barger},\ and\ \citenamefont
  {Lu}}]{Ahlers:2015moa}%
  \BibitemOpen
  \bibfield  {author} {\bibinfo {author} {\bibfnamefont {M.}~\bibnamefont
  {Ahlers}}, \bibinfo {author} {\bibfnamefont {Y.}~\bibnamefont {Bai}},
  \bibinfo {author} {\bibfnamefont {V.}~\bibnamefont {Barger}}, \ and\ \bibinfo
  {author} {\bibfnamefont {R.}~\bibnamefont {Lu}},\ }\href@noop {} {\
  (\bibinfo {year} {2015})},\ \Eprint {http://arxiv.org/abs/1505.03156}
  {arXiv:1505.03156 [hep-ph]} \BibitemShut {NoStop}%
\bibitem [{\citenamefont {Neronov}\ \emph {et~al.}(2014)\citenamefont
  {Neronov}, \citenamefont {Semikoz},\ and\ \citenamefont
  {Tchernin}}]{Neronov:2013lza}%
  \BibitemOpen
  \bibfield  {author} {\bibinfo {author} {\bibfnamefont {A.}~\bibnamefont
  {Neronov}}, \bibinfo {author} {\bibfnamefont {D.}~\bibnamefont {Semikoz}}, \
  and\ \bibinfo {author} {\bibfnamefont {C.}~\bibnamefont {Tchernin}},\ }\href
  {\doibase 10.1103/PhysRevD.89.103002} {\bibfield  {journal} {\bibinfo
  {journal} {Phys.Rev.}\ }\textbf {\bibinfo {volume} {D89}},\ \bibinfo {pages}
  {103002} (\bibinfo {year} {2014})},\ \Eprint {http://arxiv.org/abs/1307.2158}
  {arXiv:1307.2158 [astro-ph.HE]} \BibitemShut {NoStop}%
\bibitem [{\citenamefont {Ahlers}\ and\ \citenamefont
  {Murase}(2014)}]{Ahlers:2013xia}%
  \BibitemOpen
  \bibfield  {author} {\bibinfo {author} {\bibfnamefont {M.}~\bibnamefont
  {Ahlers}}\ and\ \bibinfo {author} {\bibfnamefont {K.}~\bibnamefont
  {Murase}},\ }\href {\doibase 10.1103/PhysRevD.90.023010} {\bibfield
  {journal} {\bibinfo  {journal} {Phys.Rev.}\ }\textbf {\bibinfo {volume}
  {D90}},\ \bibinfo {pages} {023010} (\bibinfo {year} {2014})},\ \Eprint
  {http://arxiv.org/abs/1309.4077} {arXiv:1309.4077 [astro-ph.HE]} \BibitemShut
  {NoStop}%
\bibitem [{\citenamefont {Gupta}(2013)}]{Gupta:2013xfa}%
  \BibitemOpen
  \bibfield  {author} {\bibinfo {author} {\bibfnamefont {N.}~\bibnamefont
  {Gupta}},\ }\href {\doibase 10.1016/j.astropartphys.2013.07.003} {\bibfield
  {journal} {\bibinfo  {journal} {Astropart.Phys.}\ }\textbf {\bibinfo {volume}
  {48}},\ \bibinfo {pages} {75} (\bibinfo {year} {2013})},\ \Eprint
  {http://arxiv.org/abs/1305.4123} {arXiv:1305.4123 [astro-ph.HE]} \BibitemShut
  {NoStop}%
\bibitem [{\citenamefont {Lunardini}\ \emph {et~al.}(2014)\citenamefont
  {Lunardini}, \citenamefont {Razzaque}, \citenamefont {Theodoseau},\ and\
  \citenamefont {Yang}}]{Lunardini:2013gva}%
  \BibitemOpen
  \bibfield  {author} {\bibinfo {author} {\bibfnamefont {C.}~\bibnamefont
  {Lunardini}}, \bibinfo {author} {\bibfnamefont {S.}~\bibnamefont {Razzaque}},
  \bibinfo {author} {\bibfnamefont {K.~T.}\ \bibnamefont {Theodoseau}}, \ and\
  \bibinfo {author} {\bibfnamefont {L.}~\bibnamefont {Yang}},\ }\href {\doibase
  10.1103/PhysRevD.90.023016} {\bibfield  {journal} {\bibinfo  {journal}
  {Phys.Rev.}\ }\textbf {\bibinfo {volume} {D90}},\ \bibinfo {pages} {023016}
  (\bibinfo {year} {2014})},\ \Eprint {http://arxiv.org/abs/1311.7188}
  {arXiv:1311.7188 [astro-ph.HE]} \BibitemShut {NoStop}%
\bibitem [{\citenamefont {Joshi}\ \emph {et~al.}(2014)\citenamefont {Joshi},
  \citenamefont {Winter},\ and\ \citenamefont {Gupta}}]{Joshi:2013aua}%
  \BibitemOpen
  \bibfield  {author} {\bibinfo {author} {\bibfnamefont {J.~C.}\ \bibnamefont
  {Joshi}}, \bibinfo {author} {\bibfnamefont {W.}~\bibnamefont {Winter}}, \
  and\ \bibinfo {author} {\bibfnamefont {N.}~\bibnamefont {Gupta}},\ }\href
  {\doibase 10.1093/mnras/stu189, 10.1093/mnras/stu2132} {\bibfield  {journal}
  {\bibinfo  {journal} {Mon.Not.Roy.Astron.Soc.}\ }\textbf {\bibinfo {volume}
  {439}},\ \bibinfo {pages} {3414} (\bibinfo {year} {2014})},\ \Eprint
  {http://arxiv.org/abs/1310.5123} {arXiv:1310.5123 [astro-ph.HE]} \BibitemShut
  {NoStop}%
\bibitem [{\citenamefont {Taylor}\ \emph {et~al.}(2014)\citenamefont {Taylor},
  \citenamefont {Gabici},\ and\ \citenamefont {Aharonian}}]{Taylor:2014hya}%
  \BibitemOpen
  \bibfield  {author} {\bibinfo {author} {\bibfnamefont {A.~M.}\ \bibnamefont
  {Taylor}}, \bibinfo {author} {\bibfnamefont {S.}~\bibnamefont {Gabici}}, \
  and\ \bibinfo {author} {\bibfnamefont {F.}~\bibnamefont {Aharonian}},\ }\href
  {\doibase 10.1103/PhysRevD.89.103003} {\bibfield  {journal} {\bibinfo
  {journal} {Phys.Rev.}\ }\textbf {\bibinfo {volume} {D89}},\ \bibinfo {pages}
  {103003} (\bibinfo {year} {2014})},\ \Eprint {http://arxiv.org/abs/1403.3206}
  {arXiv:1403.3206 [astro-ph.HE]} \BibitemShut {NoStop}%
\bibitem [{\citenamefont {Aartsen}\ \emph {et~al.}(2015)\citenamefont {Aartsen}
  \emph {et~al.}}]{Aartsen:2015ivb}%
  \BibitemOpen
  \bibfield  {author} {\bibinfo {author} {\bibfnamefont {M.}~\bibnamefont
  {Aartsen}} \emph {et~al.} (\bibinfo {collaboration} {IceCube}),\ }\href
  {\doibase 10.1103/PhysRevLett.114.171102} {\bibfield  {journal} {\bibinfo
  {journal} {Phys.Rev.Lett.}\ }\textbf {\bibinfo {volume} {114}},\ \bibinfo
  {pages} {171102} (\bibinfo {year} {2015})},\ \Eprint
  {http://arxiv.org/abs/1502.03376} {arXiv:1502.03376 [astro-ph.HE]}
  \BibitemShut {NoStop}%
\bibitem [{\citenamefont {Cholis}\ and\ \citenamefont
  {Hooper}(2013)}]{Cholis:2012kq}%
  \BibitemOpen
  \bibfield  {author} {\bibinfo {author} {\bibfnamefont {I.}~\bibnamefont
  {Cholis}}\ and\ \bibinfo {author} {\bibfnamefont {D.}~\bibnamefont
  {Hooper}},\ }\href {\doibase 10.1088/1475-7516/2013/06/030} {\bibfield
  {journal} {\bibinfo  {journal} {JCAP}\ }\textbf {\bibinfo {volume} {1306}},\
  \bibinfo {pages} {030} (\bibinfo {year} {2013})},\ \Eprint
  {http://arxiv.org/abs/1211.1974} {arXiv:1211.1974 [astro-ph.HE]} \BibitemShut
  {NoStop}%
\bibitem [{\citenamefont {Waxman}\ and\ \citenamefont
  {Bahcall}(1998)}]{Waxman:1998yy}%
  \BibitemOpen
  \bibfield  {author} {\bibinfo {author} {\bibfnamefont {E.}~\bibnamefont
  {Waxman}}\ and\ \bibinfo {author} {\bibfnamefont {J.~N.}\ \bibnamefont
  {Bahcall}},\ }\href {\doibase 10.1103/PhysRevD.59.023002} {\bibfield
  {journal} {\bibinfo  {journal} {Phys.Rev.}\ }\textbf {\bibinfo {volume}
  {D59}},\ \bibinfo {pages} {023002} (\bibinfo {year} {1998})},\ \Eprint
  {http://arxiv.org/abs/hep-ph/9807282} {arXiv:hep-ph/9807282 [hep-ph]}
  \BibitemShut {NoStop}%
\bibitem [{\citenamefont {Bahcall}\ and\ \citenamefont
  {Waxman}(2001)}]{Bahcall:1999yr}%
  \BibitemOpen
  \bibfield  {author} {\bibinfo {author} {\bibfnamefont {J.~N.}\ \bibnamefont
  {Bahcall}}\ and\ \bibinfo {author} {\bibfnamefont {E.}~\bibnamefont
  {Waxman}},\ }\href {\doibase 10.1103/PhysRevD.64.023002} {\bibfield
  {journal} {\bibinfo  {journal} {Phys.Rev.}\ }\textbf {\bibinfo {volume}
  {D64}},\ \bibinfo {pages} {023002} (\bibinfo {year} {2001})},\ \Eprint
  {http://arxiv.org/abs/hep-ph/9902383} {arXiv:hep-ph/9902383 [hep-ph]}
  \BibitemShut {NoStop}%
\bibitem [{\citenamefont {Mena}\ \emph {et~al.}(2014)\citenamefont {Mena},
  \citenamefont {Palomares-Ruiz},\ and\ \citenamefont
  {Vincent}}]{Mena:2014sja}%
  \BibitemOpen
  \bibfield  {author} {\bibinfo {author} {\bibfnamefont {O.}~\bibnamefont
  {Mena}}, \bibinfo {author} {\bibfnamefont {S.}~\bibnamefont
  {Palomares-Ruiz}}, \ and\ \bibinfo {author} {\bibfnamefont {A.~C.}\
  \bibnamefont {Vincent}},\ }\href {\doibase 10.1103/PhysRevLett.113.091103}
  {\bibfield  {journal} {\bibinfo  {journal} {Phys.Rev.Lett.}\ }\textbf
  {\bibinfo {volume} {113}},\ \bibinfo {pages} {091103} (\bibinfo {year}
  {2014})},\ \Eprint {http://arxiv.org/abs/1404.0017} {arXiv:1404.0017
  [astro-ph.HE]} \BibitemShut {NoStop}%
\bibitem [{\citenamefont {Murase}\ \emph {et~al.}(2014)\citenamefont {Murase},
  \citenamefont {Inoue},\ and\ \citenamefont {Dermer}}]{Murase:2014foa}%
  \BibitemOpen
  \bibfield  {author} {\bibinfo {author} {\bibfnamefont {K.}~\bibnamefont
  {Murase}}, \bibinfo {author} {\bibfnamefont {Y.}~\bibnamefont {Inoue}}, \
  and\ \bibinfo {author} {\bibfnamefont {C.~D.}\ \bibnamefont {Dermer}},\
  }\href {\doibase 10.1103/PhysRevD.90.023007} {\bibfield  {journal} {\bibinfo
  {journal} {Phys.Rev.}\ }\textbf {\bibinfo {volume} {D90}},\ \bibinfo {pages}
  {023007} (\bibinfo {year} {2014})},\ \Eprint {http://arxiv.org/abs/1403.4089}
  {arXiv:1403.4089 [astro-ph.HE]} \BibitemShut {NoStop}%
\bibitem [{\citenamefont {Dermer}\ \emph {et~al.}(2014)\citenamefont {Dermer},
  \citenamefont {Murase},\ and\ \citenamefont {Inoue}}]{Dermer:2014vaa}%
  \BibitemOpen
  \bibfield  {author} {\bibinfo {author} {\bibfnamefont {C.~D.}\ \bibnamefont
  {Dermer}}, \bibinfo {author} {\bibfnamefont {K.}~\bibnamefont {Murase}}, \
  and\ \bibinfo {author} {\bibfnamefont {Y.}~\bibnamefont {Inoue}},\ }\href
  {\doibase 10.1016/j.jheap.2014.09.001} {\bibfield  {journal} {\bibinfo
  {journal} {JHEAp}\ }\textbf {\bibinfo {volume} {3-4}},\ \bibinfo {pages} {29}
  (\bibinfo {year} {2014})},\ \Eprint {http://arxiv.org/abs/1406.2633}
  {arXiv:1406.2633 [astro-ph.HE]} \BibitemShut {NoStop}%
\bibitem [{\citenamefont {Stecker}(2013)}]{Stecker:2013fxa}%
  \BibitemOpen
  \bibfield  {author} {\bibinfo {author} {\bibfnamefont {F.~W.}\ \bibnamefont
  {Stecker}},\ }\href {\doibase 10.1103/PhysRevD.88.047301} {\bibfield
  {journal} {\bibinfo  {journal} {Phys.Rev.}\ }\textbf {\bibinfo {volume}
  {D88}},\ \bibinfo {pages} {047301} (\bibinfo {year} {2013})},\ \Eprint
  {http://arxiv.org/abs/1305.7404} {arXiv:1305.7404 [astro-ph.HE]} \BibitemShut
  {NoStop}%
\bibitem [{\citenamefont {Wang}\ and\ \citenamefont {Li}(2015)}]{Wang:2015woa}%
  \BibitemOpen
  \bibfield  {author} {\bibinfo {author} {\bibfnamefont {B.}~\bibnamefont
  {Wang}}\ and\ \bibinfo {author} {\bibfnamefont {Z.}~\bibnamefont {Li}},\
  }\href@noop {} {\  (\bibinfo {year} {2015})},\ \Eprint
  {http://arxiv.org/abs/1505.04418} {arXiv:1505.04418 [astro-ph.HE]}
  \BibitemShut {NoStop}%
\bibitem [{\citenamefont {Liu}\ \emph {et~al.}(2014)\citenamefont {Liu},
  \citenamefont {Wang}, \citenamefont {Inoue}, \citenamefont {Crocker},\ and\
  \citenamefont {Aharonian}}]{Liu:2013wia}%
  \BibitemOpen
  \bibfield  {author} {\bibinfo {author} {\bibfnamefont {R.-Y.}\ \bibnamefont
  {Liu}}, \bibinfo {author} {\bibfnamefont {X.-Y.}\ \bibnamefont {Wang}},
  \bibinfo {author} {\bibfnamefont {S.}~\bibnamefont {Inoue}}, \bibinfo
  {author} {\bibfnamefont {R.}~\bibnamefont {Crocker}}, \ and\ \bibinfo
  {author} {\bibfnamefont {F.}~\bibnamefont {Aharonian}},\ }\href {\doibase
  10.1103/PhysRevD.89.083004} {\bibfield  {journal} {\bibinfo  {journal}
  {Phys.Rev.}\ }\textbf {\bibinfo {volume} {D89}},\ \bibinfo {pages} {083004}
  (\bibinfo {year} {2014})},\ \Eprint {http://arxiv.org/abs/1310.1263}
  {arXiv:1310.1263 [astro-ph.HE]} \BibitemShut {NoStop}%
\bibitem [{\citenamefont {He}\ \emph {et~al.}(2013)\citenamefont {He},
  \citenamefont {Wang}, \citenamefont {Fan}, \citenamefont {Liu},\ and\
  \citenamefont {Wei}}]{He:2013cqa}%
  \BibitemOpen
  \bibfield  {author} {\bibinfo {author} {\bibfnamefont {H.-N.}\ \bibnamefont
  {He}}, \bibinfo {author} {\bibfnamefont {T.}~\bibnamefont {Wang}}, \bibinfo
  {author} {\bibfnamefont {Y.-Z.}\ \bibnamefont {Fan}}, \bibinfo {author}
  {\bibfnamefont {S.-M.}\ \bibnamefont {Liu}}, \ and\ \bibinfo {author}
  {\bibfnamefont {D.-M.}\ \bibnamefont {Wei}},\ }\href {\doibase
  10.1103/PhysRevD.87.063011} {\bibfield  {journal} {\bibinfo  {journal}
  {Phys.Rev.}\ }\textbf {\bibinfo {volume} {D87}},\ \bibinfo {pages} {063011}
  (\bibinfo {year} {2013})},\ \Eprint {http://arxiv.org/abs/1303.1253}
  {arXiv:1303.1253 [astro-ph.HE]} \BibitemShut {NoStop}%
\bibitem [{\citenamefont {Wang}\ \emph {et~al.}(2014)\citenamefont {Wang},
  \citenamefont {Zhao},\ and\ \citenamefont {Li}}]{Wang:2014jca}%
  \BibitemOpen
  \bibfield  {author} {\bibinfo {author} {\bibfnamefont {B.}~\bibnamefont
  {Wang}}, \bibinfo {author} {\bibfnamefont {X.-H.}\ \bibnamefont {Zhao}}, \
  and\ \bibinfo {author} {\bibfnamefont {Z.}~\bibnamefont {Li}},\ }\href
  {\doibase 10.1088/1475-7516/2014/11/028} {\bibfield  {journal} {\bibinfo
  {journal} {JCAP}\ }\textbf {\bibinfo {volume} {1411}},\ \bibinfo {pages}
  {028} (\bibinfo {year} {2014})},\ \Eprint {http://arxiv.org/abs/1407.2536}
  {arXiv:1407.2536 [astro-ph.HE]} \BibitemShut {NoStop}%
\bibitem [{\citenamefont {Tamborra}\ \emph {et~al.}(2014)\citenamefont
  {Tamborra}, \citenamefont {Ando},\ and\ \citenamefont
  {Murase}}]{Tamborra:2014xia}%
  \BibitemOpen
  \bibfield  {author} {\bibinfo {author} {\bibfnamefont {I.}~\bibnamefont
  {Tamborra}}, \bibinfo {author} {\bibfnamefont {S.}~\bibnamefont {Ando}}, \
  and\ \bibinfo {author} {\bibfnamefont {K.}~\bibnamefont {Murase}},\ }\href
  {\doibase 10.1088/1475-7516/2014/09/043} {\bibfield  {journal} {\bibinfo
  {journal} {JCAP}\ }\textbf {\bibinfo {volume} {1409}},\ \bibinfo {pages}
  {043} (\bibinfo {year} {2014})},\ \Eprint {http://arxiv.org/abs/1404.1189}
  {arXiv:1404.1189 [astro-ph.HE]} \BibitemShut {NoStop}%
\bibitem [{\citenamefont {Liu}\ and\ \citenamefont {Wang}(2013)}]{Liu:2012pf}%
  \BibitemOpen
  \bibfield  {author} {\bibinfo {author} {\bibfnamefont {R.-Y.}\ \bibnamefont
  {Liu}}\ and\ \bibinfo {author} {\bibfnamefont {X.-Y.}\ \bibnamefont {Wang}},\
  }\href {\doibase 10.1088/0004-637X/766/2/73} {\bibfield  {journal} {\bibinfo
  {journal} {Astrophys.J.}\ }\textbf {\bibinfo {volume} {766}},\ \bibinfo
  {pages} {73} (\bibinfo {year} {2013})},\ \Eprint
  {http://arxiv.org/abs/1212.1260} {arXiv:1212.1260 [astro-ph.HE]} \BibitemShut
  {NoStop}%
\bibitem [{\citenamefont {Murase}\ and\ \citenamefont
  {Ioka}(2013)}]{Murase:2013ffa}%
  \BibitemOpen
  \bibfield  {author} {\bibinfo {author} {\bibfnamefont {K.}~\bibnamefont
  {Murase}}\ and\ \bibinfo {author} {\bibfnamefont {K.}~\bibnamefont {Ioka}},\
  }\href {\doibase 10.1103/PhysRevLett.111.121102} {\bibfield  {journal}
  {\bibinfo  {journal} {Phys.Rev.Lett.}\ }\textbf {\bibinfo {volume} {111}},\
  \bibinfo {pages} {121102} (\bibinfo {year} {2013})},\ \Eprint
  {http://arxiv.org/abs/1306.2274} {arXiv:1306.2274 [astro-ph.HE]} \BibitemShut
  {NoStop}%
\bibitem [{\citenamefont {Xiao}\ and\ \citenamefont
  {Dai}(2015)}]{Xiao:2015gea}%
  \BibitemOpen
  \bibfield  {author} {\bibinfo {author} {\bibfnamefont {D.}~\bibnamefont
  {Xiao}}\ and\ \bibinfo {author} {\bibfnamefont {Z.}~\bibnamefont {Dai}},\
  }\href@noop {} {\  (\bibinfo {year} {2015})},\ \Eprint
  {http://arxiv.org/abs/1504.01603} {arXiv:1504.01603 [astro-ph.HE]}
  \BibitemShut {NoStop}%
\bibitem [{\citenamefont {Murase}\ \emph {et~al.}(2015)\citenamefont {Murase},
  \citenamefont {Laha}, \citenamefont {Ando},\ and\ \citenamefont
  {Ahlers}}]{Murase:2015gea}%
  \BibitemOpen
  \bibfield  {author} {\bibinfo {author} {\bibfnamefont {K.}~\bibnamefont
  {Murase}}, \bibinfo {author} {\bibfnamefont {R.}~\bibnamefont {Laha}},
  \bibinfo {author} {\bibfnamefont {S.}~\bibnamefont {Ando}}, \ and\ \bibinfo
  {author} {\bibfnamefont {M.}~\bibnamefont {Ahlers}},\ }\href@noop {} {\
  (\bibinfo {year} {2015})},\ \Eprint {http://arxiv.org/abs/1503.04663}
  {arXiv:1503.04663 [hep-ph]} \BibitemShut {NoStop}%
\bibitem [{\citenamefont {Kopp}\ \emph {et~al.}(2015)\citenamefont {Kopp},
  \citenamefont {Liu},\ and\ \citenamefont {Wang}}]{Kopp:2015bfa}%
  \BibitemOpen
  \bibfield  {author} {\bibinfo {author} {\bibfnamefont {J.}~\bibnamefont
  {Kopp}}, \bibinfo {author} {\bibfnamefont {J.}~\bibnamefont {Liu}}, \ and\
  \bibinfo {author} {\bibfnamefont {X.-P.}\ \bibnamefont {Wang}},\ }\href
  {\doibase 10.1007/JHEP04(2015)105} {\bibfield  {journal} {\bibinfo  {journal}
  {JHEP}\ }\textbf {\bibinfo {volume} {1504}},\ \bibinfo {pages} {105}
  (\bibinfo {year} {2015})},\ \Eprint {http://arxiv.org/abs/1503.02669}
  {arXiv:1503.02669 [hep-ph]} \BibitemShut {NoStop}%
\bibitem [{\citenamefont {Daikoku}\ and\ \citenamefont
  {Okada}(2015)}]{Daikoku:2015vsa}%
  \BibitemOpen
  \bibfield  {author} {\bibinfo {author} {\bibfnamefont {Y.}~\bibnamefont
  {Daikoku}}\ and\ \bibinfo {author} {\bibfnamefont {H.}~\bibnamefont
  {Okada}},\ }\href {\doibase 10.1103/PhysRevD.91.075009} {\bibfield  {journal}
  {\bibinfo  {journal} {Phys.Rev.}\ }\textbf {\bibinfo {volume} {D91}},\
  \bibinfo {pages} {075009} (\bibinfo {year} {2015})},\ \Eprint
  {http://arxiv.org/abs/1502.07032} {arXiv:1502.07032 [hep-ph]} \BibitemShut
  {NoStop}%
\bibitem [{\citenamefont {Fong}\ \emph {et~al.}(2015)\citenamefont {Fong},
  \citenamefont {Minakata}, \citenamefont {Panes},\ and\ \citenamefont
  {Funchal}}]{Fong:2014bsa}%
  \BibitemOpen
  \bibfield  {author} {\bibinfo {author} {\bibfnamefont {C.~S.}\ \bibnamefont
  {Fong}}, \bibinfo {author} {\bibfnamefont {H.}~\bibnamefont {Minakata}},
  \bibinfo {author} {\bibfnamefont {B.}~\bibnamefont {Panes}}, \ and\ \bibinfo
  {author} {\bibfnamefont {R.~Z.}\ \bibnamefont {Funchal}},\ }\href {\doibase
  10.1007/JHEP02(2015)189} {\bibfield  {journal} {\bibinfo  {journal} {JHEP}\
  }\textbf {\bibinfo {volume} {1502}},\ \bibinfo {pages} {189} (\bibinfo {year}
  {2015})},\ \Eprint {http://arxiv.org/abs/1411.5318} {arXiv:1411.5318
  [hep-ph]} \BibitemShut {NoStop}%
\bibitem [{\citenamefont {Esmaili}\ \emph {et~al.}(2014)\citenamefont
  {Esmaili}, \citenamefont {Kang},\ and\ \citenamefont
  {Serpico}}]{Esmaili:2014rma}%
  \BibitemOpen
  \bibfield  {author} {\bibinfo {author} {\bibfnamefont {A.}~\bibnamefont
  {Esmaili}}, \bibinfo {author} {\bibfnamefont {S.~K.}\ \bibnamefont {Kang}}, \
  and\ \bibinfo {author} {\bibfnamefont {P.~D.}\ \bibnamefont {Serpico}},\
  }\href {\doibase 10.1088/1475-7516/2014/12/054} {\bibfield  {journal}
  {\bibinfo  {journal} {JCAP}\ }\textbf {\bibinfo {volume} {1412}},\ \bibinfo
  {pages} {054} (\bibinfo {year} {2014})},\ \Eprint
  {http://arxiv.org/abs/1410.5979} {arXiv:1410.5979 [hep-ph]} \BibitemShut
  {NoStop}%
\bibitem [{\citenamefont {Rott}\ \emph {et~al.}(2014)\citenamefont {Rott},
  \citenamefont {Kohri},\ and\ \citenamefont {Park}}]{Rott:2014kfa}%
  \BibitemOpen
  \bibfield  {author} {\bibinfo {author} {\bibfnamefont {C.}~\bibnamefont
  {Rott}}, \bibinfo {author} {\bibfnamefont {K.}~\bibnamefont {Kohri}}, \ and\
  \bibinfo {author} {\bibfnamefont {S.~C.}\ \bibnamefont {Park}},\ }\href@noop
  {} {\  (\bibinfo {year} {2014})},\ \Eprint {http://arxiv.org/abs/1408.4575}
  {arXiv:1408.4575 [hep-ph]} \BibitemShut {NoStop}%
\bibitem [{\citenamefont {Ema}\ \emph {et~al.}(2014)\citenamefont {Ema},
  \citenamefont {Jinno},\ and\ \citenamefont {Moroi}}]{Ema:2014ufa}%
  \BibitemOpen
  \bibfield  {author} {\bibinfo {author} {\bibfnamefont {Y.}~\bibnamefont
  {Ema}}, \bibinfo {author} {\bibfnamefont {R.}~\bibnamefont {Jinno}}, \ and\
  \bibinfo {author} {\bibfnamefont {T.}~\bibnamefont {Moroi}},\ }\href
  {\doibase 10.1007/JHEP10(2014)150} {\bibfield  {journal} {\bibinfo  {journal}
  {JHEP}\ }\textbf {\bibinfo {volume} {1410}},\ \bibinfo {pages} {150}
  (\bibinfo {year} {2014})},\ \Eprint {http://arxiv.org/abs/1408.1745}
  {arXiv:1408.1745 [hep-ph]} \BibitemShut {NoStop}%
\bibitem [{\citenamefont {Bhattacharya}\ \emph {et~al.}(2015)\citenamefont
  {Bhattacharya}, \citenamefont {Gandhi},\ and\ \citenamefont
  {Gupta}}]{Bhattacharya:2014yha}%
  \BibitemOpen
  \bibfield  {author} {\bibinfo {author} {\bibfnamefont {A.}~\bibnamefont
  {Bhattacharya}}, \bibinfo {author} {\bibfnamefont {R.}~\bibnamefont
  {Gandhi}}, \ and\ \bibinfo {author} {\bibfnamefont {A.}~\bibnamefont
  {Gupta}},\ }\href {\doibase 10.1088/1475-7516/2015/03/027} {\bibfield
  {journal} {\bibinfo  {journal} {JCAP}\ }\textbf {\bibinfo {volume} {1503}},\
  \bibinfo {pages} {027} (\bibinfo {year} {2015})},\ \Eprint
  {http://arxiv.org/abs/1407.3280} {arXiv:1407.3280 [hep-ph]} \BibitemShut
  {NoStop}%
\bibitem [{\citenamefont {Zavala}(2014)}]{Zavala:2014dla}%
  \BibitemOpen
  \bibfield  {author} {\bibinfo {author} {\bibfnamefont {J.}~\bibnamefont
  {Zavala}},\ }\href {\doibase 10.1103/PhysRevD.89.123516} {\bibfield
  {journal} {\bibinfo  {journal} {Phys.Rev.}\ }\textbf {\bibinfo {volume}
  {D89}},\ \bibinfo {pages} {123516} (\bibinfo {year} {2014})},\ \Eprint
  {http://arxiv.org/abs/1404.2932} {arXiv:1404.2932 [astro-ph.HE]} \BibitemShut
  {NoStop}%
\bibitem [{\citenamefont {Boucenna}\ \emph {et~al.}(2015)\citenamefont
  {Boucenna}, \citenamefont {Chianese}, \citenamefont {Mangano}, \citenamefont
  {Miele}, \citenamefont {Morisi}, \citenamefont {Pisanti},\ and\ \citenamefont
  {Vitagliano}}]{Boucenna:2015tra}%
  \BibitemOpen
  \bibfield  {author} {\bibinfo {author} {\bibfnamefont {S.~M.}\ \bibnamefont
  {Boucenna}}, \bibinfo {author} {\bibfnamefont {M.}~\bibnamefont {Chianese}},
  \bibinfo {author} {\bibfnamefont {G.}~\bibnamefont {Mangano}}, \bibinfo
  {author} {\bibfnamefont {G.}~\bibnamefont {Miele}}, \bibinfo {author}
  {\bibfnamefont {S.}~\bibnamefont {Morisi}}, \bibinfo {author} {\bibfnamefont
  {O.}~\bibnamefont {Pisanti}}, \ and\ \bibinfo {author} {\bibfnamefont
  {E.}~\bibnamefont {Vitagliano}},\ }\href@noop {} {\  (\bibinfo {year}
  {2015})},\ \Eprint {http://arxiv.org/abs/1507.01000} {arXiv:1507.01000
  [hep-ph]} \BibitemShut {NoStop}%
\bibitem [{\citenamefont {Langacker}(2009)}]{Langacker:2008yv}%
  \BibitemOpen
  \bibfield  {author} {\bibinfo {author} {\bibfnamefont {P.}~\bibnamefont
  {Langacker}},\ }\href {\doibase 10.1103/RevModPhys.81.1199} {\bibfield
  {journal} {\bibinfo  {journal} {Rev.Mod.Phys.}\ }\textbf {\bibinfo {volume}
  {81}},\ \bibinfo {pages} {1199} (\bibinfo {year} {2009})},\ \Eprint
  {http://arxiv.org/abs/0801.1345} {arXiv:0801.1345 [hep-ph]} \BibitemShut
  {NoStop}%
\bibitem [{\citenamefont {London}\ and\ \citenamefont
  {Rosner}(1986)}]{London:1986dk}%
  \BibitemOpen
  \bibfield  {author} {\bibinfo {author} {\bibfnamefont {D.}~\bibnamefont
  {London}}\ and\ \bibinfo {author} {\bibfnamefont {J.~L.}\ \bibnamefont
  {Rosner}},\ }\href {\doibase 10.1103/PhysRevD.34.1530} {\bibfield  {journal}
  {\bibinfo  {journal} {Phys. Rev.}\ }\textbf {\bibinfo {volume} {D34}},\
  \bibinfo {pages} {1530} (\bibinfo {year} {1986})}\BibitemShut {NoStop}%
\bibitem [{\citenamefont {Hewett}\ and\ \citenamefont
  {Rizzo}(1989)}]{Hewett:1988xc}%
  \BibitemOpen
  \bibfield  {author} {\bibinfo {author} {\bibfnamefont {J.~L.}\ \bibnamefont
  {Hewett}}\ and\ \bibinfo {author} {\bibfnamefont {T.~G.}\ \bibnamefont
  {Rizzo}},\ }\href {\doibase 10.1016/0370-1573(89)90071-9} {\bibfield
  {journal} {\bibinfo  {journal} {Phys. Rept.}\ }\textbf {\bibinfo {volume}
  {183}},\ \bibinfo {pages} {193} (\bibinfo {year} {1989})}\BibitemShut
  {NoStop}%
\bibitem [{\citenamefont {Braun}\ \emph {et~al.}(2005)\citenamefont {Braun},
  \citenamefont {He}, \citenamefont {Ovrut},\ and\ \citenamefont
  {Pantev}}]{Braun:2005bw}%
  \BibitemOpen
  \bibfield  {author} {\bibinfo {author} {\bibfnamefont {V.}~\bibnamefont
  {Braun}}, \bibinfo {author} {\bibfnamefont {Y.-H.}\ \bibnamefont {He}},
  \bibinfo {author} {\bibfnamefont {B.~A.}\ \bibnamefont {Ovrut}}, \ and\
  \bibinfo {author} {\bibfnamefont {T.}~\bibnamefont {Pantev}},\ }\href
  {\doibase 10.1088/1126-6708/2005/06/039} {\bibfield  {journal} {\bibinfo
  {journal} {JHEP}\ }\textbf {\bibinfo {volume} {0506}},\ \bibinfo {pages}
  {039} (\bibinfo {year} {2005})},\ \Eprint
  {http://arxiv.org/abs/hep-th/0502155} {arXiv:hep-th/0502155 [hep-th]}
  \BibitemShut {NoStop}%
\bibitem [{\citenamefont {Cleaver}\ \emph {et~al.}(1999)\citenamefont
  {Cleaver}, \citenamefont {Cveti{\v c}}, \citenamefont {Espinosa},
  \citenamefont {Everett}, \citenamefont {Langacker} \emph
  {et~al.}}]{Cleaver:1998gc}%
  \BibitemOpen
  \bibfield  {author} {\bibinfo {author} {\bibfnamefont {G.}~\bibnamefont
  {Cleaver}}, \bibinfo {author} {\bibfnamefont {M.}~\bibnamefont {Cveti{\v
  c}}}, \bibinfo {author} {\bibfnamefont {J.}~\bibnamefont {Espinosa}},
  \bibinfo {author} {\bibfnamefont {L.}~\bibnamefont {Everett}}, \bibinfo
  {author} {\bibfnamefont {P.}~\bibnamefont {Langacker}},  \emph {et~al.},\
  }\href {\doibase 10.1103/PhysRevD.59.055005} {\bibfield  {journal} {\bibinfo
  {journal} {Phys.Rev.}\ }\textbf {\bibinfo {volume} {D59}},\ \bibinfo {pages}
  {055005} (\bibinfo {year} {1999})},\ \Eprint
  {http://arxiv.org/abs/hep-ph/9807479} {arXiv:hep-ph/9807479 [hep-ph]}
  \BibitemShut {NoStop}%
\bibitem [{\citenamefont {Coriano}\ \emph {et~al.}(2008)\citenamefont
  {Coriano}, \citenamefont {Faraggi},\ and\ \citenamefont
  {Guzzi}}]{Coriano:2007ba}%
  \BibitemOpen
  \bibfield  {author} {\bibinfo {author} {\bibfnamefont {C.}~\bibnamefont
  {Coriano}}, \bibinfo {author} {\bibfnamefont {A.~E.}\ \bibnamefont
  {Faraggi}}, \ and\ \bibinfo {author} {\bibfnamefont {M.}~\bibnamefont
  {Guzzi}},\ }\href {\doibase 10.1140/epjc/s10052-007-0469-2} {\bibfield
  {journal} {\bibinfo  {journal} {Eur.Phys.J.}\ }\textbf {\bibinfo {volume}
  {C53}},\ \bibinfo {pages} {421} (\bibinfo {year} {2008})},\ \Eprint
  {http://arxiv.org/abs/0704.1256} {arXiv:0704.1256 [hep-ph]} \BibitemShut
  {NoStop}%
\bibitem [{\citenamefont {Faraggi}\ and\ \citenamefont
  {Nanopoulos}(1991)}]{Faraggi:1990ita}%
  \BibitemOpen
  \bibfield  {author} {\bibinfo {author} {\bibfnamefont {A.~E.}\ \bibnamefont
  {Faraggi}}\ and\ \bibinfo {author} {\bibfnamefont {D.~V.}\ \bibnamefont
  {Nanopoulos}},\ }\href {\doibase 10.1142/S0217732391002621} {\bibfield
  {journal} {\bibinfo  {journal} {Mod.Phys.Lett. A}\ }\textbf {\bibinfo
  {volume} {06}},\ \bibinfo {pages} {61} (\bibinfo {year} {1991})}\BibitemShut
  {NoStop}%
\bibitem [{\citenamefont {Giedt}(2001)}]{Giedt:2000bi}%
  \BibitemOpen
  \bibfield  {author} {\bibinfo {author} {\bibfnamefont {J.}~\bibnamefont
  {Giedt}},\ }\href {\doibase 10.1006/aphy.2001.6139} {\bibfield  {journal}
  {\bibinfo  {journal} {Annals Phys.}\ }\textbf {\bibinfo {volume} {289}},\
  \bibinfo {pages} {251} (\bibinfo {year} {2001})},\ \Eprint
  {http://arxiv.org/abs/hep-th/0009104} {arXiv:hep-th/0009104 [hep-th]}
  \BibitemShut {NoStop}%
\bibitem [{\citenamefont {Lebedev}\ \emph {et~al.}(2008)\citenamefont
  {Lebedev}, \citenamefont {Nilles}, \citenamefont {Raby}, \citenamefont
  {Ramos-Sanchez}, \citenamefont {Ratz} \emph {et~al.}}]{Lebedev:2007hv}%
  \BibitemOpen
  \bibfield  {author} {\bibinfo {author} {\bibfnamefont {O.}~\bibnamefont
  {Lebedev}}, \bibinfo {author} {\bibfnamefont {H.~P.}\ \bibnamefont {Nilles}},
  \bibinfo {author} {\bibfnamefont {S.}~\bibnamefont {Raby}}, \bibinfo {author}
  {\bibfnamefont {S.}~\bibnamefont {Ramos-Sanchez}}, \bibinfo {author}
  {\bibfnamefont {M.}~\bibnamefont {Ratz}},  \emph {et~al.},\ }\href {\doibase
  10.1103/PhysRevD.77.046013} {\bibfield  {journal} {\bibinfo  {journal}
  {Phys.Rev.}\ }\textbf {\bibinfo {volume} {D77}},\ \bibinfo {pages} {046013}
  (\bibinfo {year} {2008})},\ \Eprint {http://arxiv.org/abs/0708.2691}
  {arXiv:0708.2691 [hep-th]} \BibitemShut {NoStop}%
\bibitem [{\citenamefont {Anastasopoulos}\ \emph {et~al.}(2006)\citenamefont
  {Anastasopoulos}, \citenamefont {Bianchi}, \citenamefont {Dudas},\ and\
  \citenamefont {Kiritsis}}]{Anastasopoulos:2006cz}%
  \BibitemOpen
  \bibfield  {author} {\bibinfo {author} {\bibfnamefont {P.}~\bibnamefont
  {Anastasopoulos}}, \bibinfo {author} {\bibfnamefont {M.}~\bibnamefont
  {Bianchi}}, \bibinfo {author} {\bibfnamefont {E.}~\bibnamefont {Dudas}}, \
  and\ \bibinfo {author} {\bibfnamefont {E.}~\bibnamefont {Kiritsis}},\ }\href
  {\doibase 10.1088/1126-6708/2006/11/057} {\bibfield  {journal} {\bibinfo
  {journal} {JHEP}\ }\textbf {\bibinfo {volume} {0611}},\ \bibinfo {pages}
  {057} (\bibinfo {year} {2006})},\ \Eprint
  {http://arxiv.org/abs/hep-th/0605225} {arXiv:hep-th/0605225 [hep-th]}
  \BibitemShut {NoStop}%
\bibitem [{\citenamefont {Faraggi}(1993)}]{Faraggi:1991mu}%
  \BibitemOpen
  \bibfield  {author} {\bibinfo {author} {\bibfnamefont {A.~E.}\ \bibnamefont
  {Faraggi}},\ }\href {\doibase 10.1103/PhysRevD.47.5021} {\bibfield  {journal}
  {\bibinfo  {journal} {Phys.Rev.}\ }\textbf {\bibinfo {volume} {D47}},\
  \bibinfo {pages} {5021} (\bibinfo {year} {1993})}\BibitemShut {NoStop}%
\bibitem [{\citenamefont {Cveti{\v c}}\ \emph {et~al.}(2001)\citenamefont
  {Cveti{\v c}}, \citenamefont {Shiu},\ and\ \citenamefont
  {Uranga}}]{Cvetic:2001nr}%
  \BibitemOpen
  \bibfield  {author} {\bibinfo {author} {\bibfnamefont {M.}~\bibnamefont
  {Cveti{\v c}}}, \bibinfo {author} {\bibfnamefont {G.}~\bibnamefont {Shiu}}, \
  and\ \bibinfo {author} {\bibfnamefont {A.~M.}\ \bibnamefont {Uranga}},\
  }\href {\doibase 10.1016/S0550-3213(01)00427-8} {\bibfield  {journal}
  {\bibinfo  {journal} {Nucl.Phys.}\ }\textbf {\bibinfo {volume} {B615}},\
  \bibinfo {pages} {3} (\bibinfo {year} {2001})},\ \Eprint
  {http://arxiv.org/abs/hep-th/0107166} {arXiv:hep-th/0107166 [hep-th]}
  \BibitemShut {NoStop}%
\bibitem [{\citenamefont {Arkani-Hamed}\ \emph
  {et~al.}(2001{\natexlab{a}})\citenamefont {Arkani-Hamed}, \citenamefont
  {Cohen},\ and\ \citenamefont {Georgi}}]{ArkaniHamed:2001is}%
  \BibitemOpen
  \bibfield  {author} {\bibinfo {author} {\bibfnamefont {N.}~\bibnamefont
  {Arkani-Hamed}}, \bibinfo {author} {\bibfnamefont {A.~G.}\ \bibnamefont
  {Cohen}}, \ and\ \bibinfo {author} {\bibfnamefont {H.}~\bibnamefont
  {Georgi}},\ }\href {\doibase 10.1016/S0370-2693(01)00946-7} {\bibfield
  {journal} {\bibinfo  {journal} {Phys.Lett.}\ }\textbf {\bibinfo {volume}
  {B516}},\ \bibinfo {pages} {395} (\bibinfo {year} {2001}{\natexlab{a}})},\
  \Eprint {http://arxiv.org/abs/hep-th/0103135} {arXiv:hep-th/0103135 [hep-th]}
  \BibitemShut {NoStop}%
\bibitem [{\citenamefont {Arkani-Hamed}\ \emph {et~al.}(2002)\citenamefont
  {Arkani-Hamed} \emph {et~al.}}]{ArkaniHamed:2002qx}%
  \BibitemOpen
  \bibfield  {author} {\bibinfo {author} {\bibfnamefont {N.}~\bibnamefont
  {Arkani-Hamed}} \emph {et~al.},\ }\href@noop {} {\bibfield  {journal}
  {\bibinfo  {journal} {JHEP}\ }\textbf {\bibinfo {volume} {08}},\ \bibinfo
  {pages} {021} (\bibinfo {year} {2002})},\ \Eprint
  {http://arxiv.org/abs/hep-ph/0206020} {arXiv:hep-ph/0206020} \BibitemShut
  {NoStop}%
\bibitem [{\citenamefont {Han}\ \emph {et~al.}(2003)\citenamefont {Han},
  \citenamefont {Logan}, \citenamefont {McElrath},\ and\ \citenamefont
  {Wang}}]{Han:2003wu}%
  \BibitemOpen
  \bibfield  {author} {\bibinfo {author} {\bibfnamefont {T.}~\bibnamefont
  {Han}}, \bibinfo {author} {\bibfnamefont {H.~E.}\ \bibnamefont {Logan}},
  \bibinfo {author} {\bibfnamefont {B.}~\bibnamefont {McElrath}}, \ and\
  \bibinfo {author} {\bibfnamefont {L.-T.}\ \bibnamefont {Wang}},\ }\href
  {\doibase 10.1103/PhysRevD.67.095004} {\bibfield  {journal} {\bibinfo
  {journal} {Phys.Rev.}\ }\textbf {\bibinfo {volume} {D67}},\ \bibinfo {pages}
  {095004} (\bibinfo {year} {2003})},\ \Eprint
  {http://arxiv.org/abs/hep-ph/0301040} {arXiv:hep-ph/0301040 [hep-ph]}
  \BibitemShut {NoStop}%
\bibitem [{\citenamefont {Perelstein}(2007)}]{Perelstein:2005ka}%
  \BibitemOpen
  \bibfield  {author} {\bibinfo {author} {\bibfnamefont {M.}~\bibnamefont
  {Perelstein}},\ }\href {\doibase 10.1016/j.ppnp.2006.04.001} {\bibfield
  {journal} {\bibinfo  {journal} {Prog.Part.Nucl.Phys.}\ }\textbf {\bibinfo
  {volume} {58}},\ \bibinfo {pages} {247} (\bibinfo {year} {2007})},\ \Eprint
  {http://arxiv.org/abs/hep-ph/0512128} {arXiv:hep-ph/0512128 [hep-ph]}
  \BibitemShut {NoStop}%
\bibitem [{\citenamefont {Hill}\ and\ \citenamefont
  {Simmons}(2003)}]{Hill:2002ap}%
  \BibitemOpen
  \bibfield  {author} {\bibinfo {author} {\bibfnamefont {C.~T.}\ \bibnamefont
  {Hill}}\ and\ \bibinfo {author} {\bibfnamefont {E.~H.}\ \bibnamefont
  {Simmons}},\ }\href {\doibase 10.1016/S0370-1573(03)00140-6} {\bibfield
  {journal} {\bibinfo  {journal} {Phys.Rept.}\ }\textbf {\bibinfo {volume}
  {381}},\ \bibinfo {pages} {235} (\bibinfo {year} {2003})},\ \Eprint
  {http://arxiv.org/abs/hep-ph/0203079} {arXiv:hep-ph/0203079 [hep-ph]}
  \BibitemShut {NoStop}%
\bibitem [{\citenamefont {Chivukula}\ \emph {et~al.}(2004)\citenamefont
  {Chivukula}, \citenamefont {He}, \citenamefont {Howard},\ and\ \citenamefont
  {Simmons}}]{Chivukula:2003wj}%
  \BibitemOpen
  \bibfield  {author} {\bibinfo {author} {\bibfnamefont {R.~S.}\ \bibnamefont
  {Chivukula}}, \bibinfo {author} {\bibfnamefont {H.-J.}\ \bibnamefont {He}},
  \bibinfo {author} {\bibfnamefont {J.}~\bibnamefont {Howard}}, \ and\ \bibinfo
  {author} {\bibfnamefont {E.~H.}\ \bibnamefont {Simmons}},\ }\href {\doibase
  10.1103/PhysRevD.69.015009} {\bibfield  {journal} {\bibinfo  {journal}
  {Phys.Rev.}\ }\textbf {\bibinfo {volume} {D69}},\ \bibinfo {pages} {015009}
  (\bibinfo {year} {2004})},\ \Eprint {http://arxiv.org/abs/hep-ph/0307209}
  {arXiv:hep-ph/0307209 [hep-ph]} \BibitemShut {NoStop}%
\bibitem [{\citenamefont {Chivukula}\ and\ \citenamefont
  {Simmons}(2002)}]{Chivukula:2002ry}%
  \BibitemOpen
  \bibfield  {author} {\bibinfo {author} {\bibfnamefont {R.}~\bibnamefont
  {Chivukula}}\ and\ \bibinfo {author} {\bibfnamefont {E.~H.}\ \bibnamefont
  {Simmons}},\ }\href {\doibase 10.1103/PhysRevD.66.015006} {\bibfield
  {journal} {\bibinfo  {journal} {Phys.Rev.}\ }\textbf {\bibinfo {volume}
  {D66}},\ \bibinfo {pages} {015006} (\bibinfo {year} {2002})},\ \Eprint
  {http://arxiv.org/abs/hep-ph/0205064} {arXiv:hep-ph/0205064 [hep-ph]}
  \BibitemShut {NoStop}%
\bibitem [{\citenamefont {Agashe}\ \emph {et~al.}(2003)\citenamefont {Agashe},
  \citenamefont {Delgado}, \citenamefont {May},\ and\ \citenamefont
  {Sundrum}}]{Agashe:2003zs}%
  \BibitemOpen
  \bibfield  {author} {\bibinfo {author} {\bibfnamefont {K.}~\bibnamefont
  {Agashe}}, \bibinfo {author} {\bibfnamefont {A.}~\bibnamefont {Delgado}},
  \bibinfo {author} {\bibfnamefont {M.~J.}\ \bibnamefont {May}}, \ and\
  \bibinfo {author} {\bibfnamefont {R.}~\bibnamefont {Sundrum}},\ }\href
  {\doibase 10.1088/1126-6708/2003/08/050} {\bibfield  {journal} {\bibinfo
  {journal} {JHEP}\ }\textbf {\bibinfo {volume} {0308}},\ \bibinfo {pages}
  {050} (\bibinfo {year} {2003})},\ \Eprint
  {http://arxiv.org/abs/hep-ph/0308036} {arXiv:hep-ph/0308036 [hep-ph]}
  \BibitemShut {NoStop}%
\bibitem [{\citenamefont {Agashe}\ \emph {et~al.}(2007)\citenamefont {Agashe},
  \citenamefont {Davoudiasl}, \citenamefont {Gopalakrishna}, \citenamefont
  {Han}, \citenamefont {Huang} \emph {et~al.}}]{Agashe:2007ki}%
  \BibitemOpen
  \bibfield  {author} {\bibinfo {author} {\bibfnamefont {K.}~\bibnamefont
  {Agashe}}, \bibinfo {author} {\bibfnamefont {H.}~\bibnamefont {Davoudiasl}},
  \bibinfo {author} {\bibfnamefont {S.}~\bibnamefont {Gopalakrishna}}, \bibinfo
  {author} {\bibfnamefont {T.}~\bibnamefont {Han}}, \bibinfo {author}
  {\bibfnamefont {G.-Y.}\ \bibnamefont {Huang}},  \emph {et~al.},\ }\href
  {\doibase 10.1103/PhysRevD.76.115015} {\bibfield  {journal} {\bibinfo
  {journal} {Phys.Rev.}\ }\textbf {\bibinfo {volume} {D76}},\ \bibinfo {pages}
  {115015} (\bibinfo {year} {2007})},\ \Eprint {http://arxiv.org/abs/0709.0007}
  {arXiv:0709.0007 [hep-ph]} \BibitemShut {NoStop}%
\bibitem [{\citenamefont {Carena}\ \emph {et~al.}(2003)\citenamefont {Carena},
  \citenamefont {Delgado}, \citenamefont {Ponton}, \citenamefont {Tait},\ and\
  \citenamefont {Wagner}}]{Carena:2003fx}%
  \BibitemOpen
  \bibfield  {author} {\bibinfo {author} {\bibfnamefont {M.~S.}\ \bibnamefont
  {Carena}}, \bibinfo {author} {\bibfnamefont {A.}~\bibnamefont {Delgado}},
  \bibinfo {author} {\bibfnamefont {E.}~\bibnamefont {Ponton}}, \bibinfo
  {author} {\bibfnamefont {T.~M.}\ \bibnamefont {Tait}}, \ and\ \bibinfo
  {author} {\bibfnamefont {C.}~\bibnamefont {Wagner}},\ }\href {\doibase
  10.1103/PhysRevD.68.035010} {\bibfield  {journal} {\bibinfo  {journal}
  {Phys.Rev.}\ }\textbf {\bibinfo {volume} {D68}},\ \bibinfo {pages} {035010}
  (\bibinfo {year} {2003})},\ \Eprint {http://arxiv.org/abs/hep-ph/0305188}
  {arXiv:hep-ph/0305188 [hep-ph]} \BibitemShut {NoStop}%
\bibitem [{\citenamefont {Hewett}\ \emph {et~al.}(2002)\citenamefont {Hewett},
  \citenamefont {Petriello},\ and\ \citenamefont {Rizzo}}]{Hewett:2002fe}%
  \BibitemOpen
  \bibfield  {author} {\bibinfo {author} {\bibfnamefont {J.}~\bibnamefont
  {Hewett}}, \bibinfo {author} {\bibfnamefont {F.}~\bibnamefont {Petriello}}, \
  and\ \bibinfo {author} {\bibfnamefont {T.}~\bibnamefont {Rizzo}},\ }\href
  {\doibase 10.1088/1126-6708/2002/09/030} {\bibfield  {journal} {\bibinfo
  {journal} {JHEP}\ }\textbf {\bibinfo {volume} {0209}},\ \bibinfo {pages}
  {030} (\bibinfo {year} {2002})},\ \Eprint
  {http://arxiv.org/abs/hep-ph/0203091} {arXiv:hep-ph/0203091 [hep-ph]}
  \BibitemShut {NoStop}%
\bibitem [{\citenamefont {Arkani-Hamed}\ \emph
  {et~al.}(2001{\natexlab{b}})\citenamefont {Arkani-Hamed}, \citenamefont
  {Cohen},\ and\ \citenamefont {Georgi}}]{ArkaniHamed:2001nc}%
  \BibitemOpen
  \bibfield  {author} {\bibinfo {author} {\bibfnamefont {N.}~\bibnamefont
  {Arkani-Hamed}}, \bibinfo {author} {\bibfnamefont {A.~G.}\ \bibnamefont
  {Cohen}}, \ and\ \bibinfo {author} {\bibfnamefont {H.}~\bibnamefont
  {Georgi}},\ }\href {\doibase 10.1016/S0370-2693(01)00741-9} {\bibfield
  {journal} {\bibinfo  {journal} {Phys. Lett.}\ }\textbf {\bibinfo {volume}
  {B513}},\ \bibinfo {pages} {232} (\bibinfo {year} {2001}{\natexlab{b}})},\
  \Eprint {http://arxiv.org/abs/hep-ph/0105239} {arXiv:hep-ph/0105239}
  \BibitemShut {NoStop}%
\bibitem [{\citenamefont {Cveti{\v c}}\ \emph {et~al.}(1997)\citenamefont
  {Cveti{\v c}}, \citenamefont {Demir}, \citenamefont {Espinosa}, \citenamefont
  {Everett},\ and\ \citenamefont {Langacker}}]{Cvetic:1997ky}%
  \BibitemOpen
  \bibfield  {author} {\bibinfo {author} {\bibfnamefont {M.}~\bibnamefont
  {Cveti{\v c}}}, \bibinfo {author} {\bibfnamefont {D.~A.}\ \bibnamefont
  {Demir}}, \bibinfo {author} {\bibfnamefont {J.~R.}\ \bibnamefont {Espinosa}},
  \bibinfo {author} {\bibfnamefont {L.~L.}\ \bibnamefont {Everett}}, \ and\
  \bibinfo {author} {\bibfnamefont {P.}~\bibnamefont {Langacker}},\ }\href
  {\doibase 10.1103/PhysRevD.56.2861} {\bibfield  {journal} {\bibinfo
  {journal} {Phys. Rev.}\ }\textbf {\bibinfo {volume} {D56}},\ \bibinfo {pages}
  {2861} (\bibinfo {year} {1997})},\ \Eprint
  {http://arxiv.org/abs/hep-ph/9703317} {arXiv:hep-ph/9703317} \BibitemShut
  {NoStop}%
\bibitem [{\citenamefont {Langacker}\ \emph {et~al.}(1999)\citenamefont
  {Langacker}, \citenamefont {Polonsky},\ and\ \citenamefont
  {Wang}}]{Langacker:1999hs}%
  \BibitemOpen
  \bibfield  {author} {\bibinfo {author} {\bibfnamefont {P.}~\bibnamefont
  {Langacker}}, \bibinfo {author} {\bibfnamefont {N.}~\bibnamefont {Polonsky}},
  \ and\ \bibinfo {author} {\bibfnamefont {J.}~\bibnamefont {Wang}},\ }\href
  {\doibase 10.1103/PhysRevD.60.115005} {\bibfield  {journal} {\bibinfo
  {journal} {Phys. Rev.}\ }\textbf {\bibinfo {volume} {D60}},\ \bibinfo {pages}
  {115005} (\bibinfo {year} {1999})},\ \Eprint
  {http://arxiv.org/abs/hep-ph/9905252} {arXiv:hep-ph/9905252} \BibitemShut
  {NoStop}%
\bibitem [{\citenamefont {Beringer}\ \emph {et~al.}(2012)\citenamefont
  {Beringer} \emph {et~al.}}]{Beringer:1900zz}%
  \BibitemOpen
  \bibfield  {author} {\bibinfo {author} {\bibfnamefont {J.}~\bibnamefont
  {Beringer}} \emph {et~al.} (\bibinfo {collaboration} {Particle Data Group}),\
  }\href {\doibase 10.1103/PhysRevD.86.010001} {\bibfield  {journal} {\bibinfo
  {journal} {Phys.Rev.}\ }\textbf {\bibinfo {volume} {D86}},\ \bibinfo {pages}
  {010001} (\bibinfo {year} {2012})}\BibitemShut {NoStop}%
\bibitem [{\citenamefont {Carey}\ \emph {et~al.}(2009)\citenamefont {Carey},
  \citenamefont {Lynch}, \citenamefont {Miller}, \citenamefont {Roberts},
  \citenamefont {Morse} \emph {et~al.}}]{Carey:2009zzb}%
  \BibitemOpen
  \bibfield  {author} {\bibinfo {author} {\bibfnamefont {R.}~\bibnamefont
  {Carey}}, \bibinfo {author} {\bibfnamefont {K.}~\bibnamefont {Lynch}},
  \bibinfo {author} {\bibfnamefont {J.}~\bibnamefont {Miller}}, \bibinfo
  {author} {\bibfnamefont {B.}~\bibnamefont {Roberts}}, \bibinfo {author}
  {\bibfnamefont {W.}~\bibnamefont {Morse}},  \emph {et~al.},\ }\href@noop {}
  {\  (\bibinfo {year} {2009})}\BibitemShut {NoStop}%
\bibitem [{\citenamefont {Kronfeld}\ \emph {et~al.}(2013)\citenamefont
  {Kronfeld}, \citenamefont {Tschirhart}, \citenamefont {Al-Binni},
  \citenamefont {Altmannshofer}, \citenamefont {Ankenbrandt} \emph
  {et~al.}}]{Kronfeld:2013uoa}%
  \BibitemOpen
  \bibfield  {author} {\bibinfo {author} {\bibfnamefont {A.~S.}\ \bibnamefont
  {Kronfeld}}, \bibinfo {author} {\bibfnamefont {R.~S.}\ \bibnamefont
  {Tschirhart}}, \bibinfo {author} {\bibfnamefont {U.}~\bibnamefont
  {Al-Binni}}, \bibinfo {author} {\bibfnamefont {W.}~\bibnamefont
  {Altmannshofer}}, \bibinfo {author} {\bibfnamefont {C.}~\bibnamefont
  {Ankenbrandt}},  \emph {et~al.},\ }\href@noop {} {\  (\bibinfo {year}
  {2013})},\ \Eprint {http://arxiv.org/abs/1306.5009} {arXiv:1306.5009
  [hep-ex]} \BibitemShut {NoStop}%
\bibitem [{\citenamefont {Aubin}\ \emph {et~al.}(2012)\citenamefont {Aubin},
  \citenamefont {Blum}, \citenamefont {Golterman},\ and\ \citenamefont
  {Peris}}]{Aubin:2012me}%
  \BibitemOpen
  \bibfield  {author} {\bibinfo {author} {\bibfnamefont {C.}~\bibnamefont
  {Aubin}}, \bibinfo {author} {\bibfnamefont {T.}~\bibnamefont {Blum}},
  \bibinfo {author} {\bibfnamefont {M.}~\bibnamefont {Golterman}}, \ and\
  \bibinfo {author} {\bibfnamefont {S.}~\bibnamefont {Peris}},\ }\href
  {\doibase 10.1103/PhysRevD.86.054509} {\bibfield  {journal} {\bibinfo
  {journal} {Phys.Rev.}\ }\textbf {\bibinfo {volume} {D86}},\ \bibinfo {pages}
  {054509} (\bibinfo {year} {2012})},\ \Eprint {http://arxiv.org/abs/1205.3695}
  {arXiv:1205.3695 [hep-lat]} \BibitemShut {NoStop}%
\bibitem [{\citenamefont {Aubin}\ \emph {et~al.}(2013)\citenamefont {Aubin},
  \citenamefont {Blum}, \citenamefont {Golterman},\ and\ \citenamefont
  {Peris}}]{Aubin:2013daa}%
  \BibitemOpen
  \bibfield  {author} {\bibinfo {author} {\bibfnamefont {C.}~\bibnamefont
  {Aubin}}, \bibinfo {author} {\bibfnamefont {T.}~\bibnamefont {Blum}},
  \bibinfo {author} {\bibfnamefont {M.}~\bibnamefont {Golterman}}, \ and\
  \bibinfo {author} {\bibfnamefont {S.}~\bibnamefont {Peris}},\ }\href
  {\doibase 10.1103/PhysRevD.88.074505} {\bibfield  {journal} {\bibinfo
  {journal} {Phys.Rev.}\ }\textbf {\bibinfo {volume} {D88}},\ \bibinfo {pages}
  {074505} (\bibinfo {year} {2013})},\ \Eprint {http://arxiv.org/abs/1307.4701}
  {arXiv:1307.4701 [hep-lat]} \BibitemShut {NoStop}%
\bibitem [{\citenamefont {Aubin}\ \emph {et~al.}(2014)\citenamefont {Aubin},
  \citenamefont {Blum}, \citenamefont {Golterman}, \citenamefont {Maltman},\
  and\ \citenamefont {Peris}}]{Aubin:2013yba}%
  \BibitemOpen
  \bibfield  {author} {\bibinfo {author} {\bibfnamefont {C.}~\bibnamefont
  {Aubin}}, \bibinfo {author} {\bibfnamefont {T.}~\bibnamefont {Blum}},
  \bibinfo {author} {\bibfnamefont {M.}~\bibnamefont {Golterman}}, \bibinfo
  {author} {\bibfnamefont {K.}~\bibnamefont {Maltman}}, \ and\ \bibinfo
  {author} {\bibfnamefont {S.}~\bibnamefont {Peris}},\ }\href {\doibase
  10.1142/S2010194514604189} {\bibfield  {journal} {\bibinfo  {journal}
  {Int.J.Mod.Phys.Conf.Ser.}\ }\textbf {\bibinfo {volume} {35}},\ \bibinfo
  {pages} {1460418} (\bibinfo {year} {2014})},\ \Eprint
  {http://arxiv.org/abs/1311.5504} {arXiv:1311.5504 [hep-lat]} \BibitemShut
  {NoStop}%
\bibitem [{\citenamefont {Blum}\ \emph {et~al.}(2013)\citenamefont {Blum},
  \citenamefont {Denig}, \citenamefont {Logashenko}, \citenamefont {de~Rafael},
  \citenamefont {Lee~Roberts} \emph {et~al.}}]{Blum:2013xva}%
  \BibitemOpen
  \bibfield  {author} {\bibinfo {author} {\bibfnamefont {T.}~\bibnamefont
  {Blum}}, \bibinfo {author} {\bibfnamefont {A.}~\bibnamefont {Denig}},
  \bibinfo {author} {\bibfnamefont {I.}~\bibnamefont {Logashenko}}, \bibinfo
  {author} {\bibfnamefont {E.}~\bibnamefont {de~Rafael}}, \bibinfo {author}
  {\bibfnamefont {B.}~\bibnamefont {Lee~Roberts}},  \emph {et~al.},\
  }\href@noop {} {\  (\bibinfo {year} {2013})},\ \Eprint
  {http://arxiv.org/abs/1311.2198} {arXiv:1311.2198 [hep-ph]} \BibitemShut
  {NoStop}%
\bibitem [{\citenamefont {Golterman}\ \emph {et~al.}(2014)\citenamefont
  {Golterman}, \citenamefont {Maltman},\ and\ \citenamefont
  {Peris}}]{Golterman:2013vna}%
  \BibitemOpen
  \bibfield  {author} {\bibinfo {author} {\bibfnamefont {M.}~\bibnamefont
  {Golterman}}, \bibinfo {author} {\bibfnamefont {K.}~\bibnamefont {Maltman}},
  \ and\ \bibinfo {author} {\bibfnamefont {S.}~\bibnamefont {Peris}},\
  }\href@noop {} {\bibfield  {journal} {\bibinfo  {journal} {PoS}\ }\textbf
  {\bibinfo {volume} {LATTICE2013}},\ \bibinfo {pages} {300} (\bibinfo {year}
  {2014})},\ \Eprint {http://arxiv.org/abs/1310.5928} {arXiv:1310.5928
  [hep-lat]} \BibitemShut {NoStop}%
\bibitem [{\citenamefont {Nyffeler}(2014)}]{Nyffeler:2013lia}%
  \BibitemOpen
  \bibfield  {author} {\bibinfo {author} {\bibfnamefont {A.}~\bibnamefont
  {Nyffeler}},\ }\href {\doibase 10.1393/ncc/i2014-11752-0} {\bibfield
  {journal} {\bibinfo  {journal} {Nuovo Cim.}\ }\textbf {\bibinfo {volume}
  {C037}},\ \bibinfo {pages} {173} (\bibinfo {year} {2014})},\ \Eprint
  {http://arxiv.org/abs/1312.4804} {arXiv:1312.4804 [hep-ph]} \BibitemShut
  {NoStop}%
\bibitem [{\citenamefont {Endo}\ \emph {et~al.}(2013)\citenamefont {Endo},
  \citenamefont {Hamaguchi}, \citenamefont {Kitahara},\ and\ \citenamefont
  {Yoshinaga}}]{Endo:2013lva}%
  \BibitemOpen
  \bibfield  {author} {\bibinfo {author} {\bibfnamefont {M.}~\bibnamefont
  {Endo}}, \bibinfo {author} {\bibfnamefont {K.}~\bibnamefont {Hamaguchi}},
  \bibinfo {author} {\bibfnamefont {T.}~\bibnamefont {Kitahara}}, \ and\
  \bibinfo {author} {\bibfnamefont {T.}~\bibnamefont {Yoshinaga}},\ }\href
  {\doibase 10.1007/JHEP11(2013)013} {\bibfield  {journal} {\bibinfo  {journal}
  {JHEP}\ }\textbf {\bibinfo {volume} {1311}},\ \bibinfo {pages} {013}
  (\bibinfo {year} {2013})},\ \Eprint {http://arxiv.org/abs/1309.3065}
  {arXiv:1309.3065 [hep-ph]} \BibitemShut {NoStop}%
\bibitem [{\citenamefont {Ibe}\ \emph {et~al.}(2013)\citenamefont {Ibe},
  \citenamefont {Yanagida},\ and\ \citenamefont {Yokozaki}}]{Ibe:2013oha}%
  \BibitemOpen
  \bibfield  {author} {\bibinfo {author} {\bibfnamefont {M.}~\bibnamefont
  {Ibe}}, \bibinfo {author} {\bibfnamefont {T.~T.}\ \bibnamefont {Yanagida}}, \
  and\ \bibinfo {author} {\bibfnamefont {N.}~\bibnamefont {Yokozaki}},\ }\href
  {\doibase 10.1007/JHEP08(2013)067} {\bibfield  {journal} {\bibinfo  {journal}
  {JHEP}\ }\textbf {\bibinfo {volume} {1308}},\ \bibinfo {pages} {067}
  (\bibinfo {year} {2013})},\ \Eprint {http://arxiv.org/abs/1303.6995}
  {arXiv:1303.6995 [hep-ph]} \BibitemShut {NoStop}%
\bibitem [{\citenamefont {Davoudiasl}\ \emph {et~al.}(2014)\citenamefont
  {Davoudiasl}, \citenamefont {Lee},\ and\ \citenamefont
  {Marciano}}]{Davoudiasl:2014kua}%
  \BibitemOpen
  \bibfield  {author} {\bibinfo {author} {\bibfnamefont {H.}~\bibnamefont
  {Davoudiasl}}, \bibinfo {author} {\bibfnamefont {H.-S.}\ \bibnamefont {Lee}},
  \ and\ \bibinfo {author} {\bibfnamefont {W.~J.}\ \bibnamefont {Marciano}},\
  }\href {\doibase 10.1103/PhysRevD.89.095006} {\bibfield  {journal} {\bibinfo
  {journal} {Phys.Rev.}\ }\textbf {\bibinfo {volume} {D89}},\ \bibinfo {pages}
  {095006} (\bibinfo {year} {2014})},\ \Eprint {http://arxiv.org/abs/1402.3620}
  {arXiv:1402.3620 [hep-ph]} \BibitemShut {NoStop}%
\bibitem [{\citenamefont {Agrawal}\ \emph {et~al.}(2014)\citenamefont
  {Agrawal}, \citenamefont {Chacko},\ and\ \citenamefont
  {Verhaaren}}]{Agrawal:2014ufa}%
  \BibitemOpen
  \bibfield  {author} {\bibinfo {author} {\bibfnamefont {P.}~\bibnamefont
  {Agrawal}}, \bibinfo {author} {\bibfnamefont {Z.}~\bibnamefont {Chacko}}, \
  and\ \bibinfo {author} {\bibfnamefont {C.~B.}\ \bibnamefont {Verhaaren}},\
  }\href {\doibase 10.1007/JHEP08(2014)147} {\bibfield  {journal} {\bibinfo
  {journal} {JHEP}\ }\textbf {\bibinfo {volume} {1408}},\ \bibinfo {pages}
  {147} (\bibinfo {year} {2014})},\ \Eprint {http://arxiv.org/abs/1402.7369}
  {arXiv:1402.7369 [hep-ph]} \BibitemShut {NoStop}%
\bibitem [{\citenamefont {Ajaib}\ \emph {et~al.}(2014)\citenamefont {Ajaib},
  \citenamefont {Gogoladze}, \citenamefont {Shafi},\ and\ \citenamefont
  {Ün}}]{Ajaib:2014ana}%
  \BibitemOpen
  \bibfield  {author} {\bibinfo {author} {\bibfnamefont {M.~A.}\ \bibnamefont
  {Ajaib}}, \bibinfo {author} {\bibfnamefont {I.}~\bibnamefont {Gogoladze}},
  \bibinfo {author} {\bibfnamefont {Q.}~\bibnamefont {Shafi}}, \ and\ \bibinfo
  {author} {\bibfnamefont {C.~S.}\ \bibnamefont {Ün}},\ }\href {\doibase
  10.1007/JHEP05(2014)079} {\bibfield  {journal} {\bibinfo  {journal} {JHEP}\
  }\textbf {\bibinfo {volume} {1405}},\ \bibinfo {pages} {079} (\bibinfo {year}
  {2014})},\ \Eprint {http://arxiv.org/abs/1402.4918} {arXiv:1402.4918
  [hep-ph]} \BibitemShut {NoStop}%
\bibitem [{\citenamefont {McKeen}(2011)}]{McKeen:2009ny}%
  \BibitemOpen
  \bibfield  {author} {\bibinfo {author} {\bibfnamefont {D.}~\bibnamefont
  {McKeen}},\ }\href {\doibase 10.1016/j.aop.2010.12.015} {\bibfield  {journal}
  {\bibinfo  {journal} {Annals Phys.}\ }\textbf {\bibinfo {volume} {326}},\
  \bibinfo {pages} {1501} (\bibinfo {year} {2011})},\ \Eprint
  {http://arxiv.org/abs/0912.1076} {arXiv:0912.1076 [hep-ph]} \BibitemShut
  {NoStop}%
\bibitem [{\citenamefont {Jegerlehner}\ and\ \citenamefont
  {Nyffeler}(2009)}]{Jegerlehner:2009ry}%
  \BibitemOpen
  \bibfield  {author} {\bibinfo {author} {\bibfnamefont {F.}~\bibnamefont
  {Jegerlehner}}\ and\ \bibinfo {author} {\bibfnamefont {A.}~\bibnamefont
  {Nyffeler}},\ }\href {\doibase 10.1016/j.physrep.2009.04.003} {\bibfield
  {journal} {\bibinfo  {journal} {Phys.Rept.}\ }\textbf {\bibinfo {volume}
  {477}},\ \bibinfo {pages} {1} (\bibinfo {year} {2009})},\ \Eprint
  {http://arxiv.org/abs/0902.3360} {arXiv:0902.3360 [hep-ph]} \BibitemShut
  {NoStop}%
\bibitem [{\citenamefont {Queiroz}\ and\ \citenamefont
  {Shepherd}(2014)}]{Queiroz:2014zfa}%
  \BibitemOpen
  \bibfield  {author} {\bibinfo {author} {\bibfnamefont {F.~S.}\ \bibnamefont
  {Queiroz}}\ and\ \bibinfo {author} {\bibfnamefont {W.}~\bibnamefont
  {Shepherd}},\ }\href {\doibase 10.1103/PhysRevD.89.095024} {\bibfield
  {journal} {\bibinfo  {journal} {Phys.Rev.}\ }\textbf {\bibinfo {volume}
  {D89}},\ \bibinfo {pages} {095024} (\bibinfo {year} {2014})},\ \Eprint
  {http://arxiv.org/abs/1403.2309} {arXiv:1403.2309 [hep-ph]} \BibitemShut
  {NoStop}%
\bibitem [{\citenamefont {Altmannshofer}\ \emph {et~al.}(2014)\citenamefont
  {Altmannshofer}, \citenamefont {Gori}, \citenamefont {Pospelov},\ and\
  \citenamefont {Yavin}}]{Altmannshofer:2014pba}%
  \BibitemOpen
  \bibfield  {author} {\bibinfo {author} {\bibfnamefont {W.}~\bibnamefont
  {Altmannshofer}}, \bibinfo {author} {\bibfnamefont {S.}~\bibnamefont {Gori}},
  \bibinfo {author} {\bibfnamefont {M.}~\bibnamefont {Pospelov}}, \ and\
  \bibinfo {author} {\bibfnamefont {I.}~\bibnamefont {Yavin}},\ }\href
  {\doibase 10.1103/PhysRevLett.113.091801} {\bibfield  {journal} {\bibinfo
  {journal} {Phys.Rev.Lett.}\ }\textbf {\bibinfo {volume} {113}},\ \bibinfo
  {pages} {091801} (\bibinfo {year} {2014})},\ \Eprint
  {http://arxiv.org/abs/1406.2332} {arXiv:1406.2332 [hep-ph]} \BibitemShut
  {NoStop}%
\bibitem [{\citenamefont {Ahlgren}\ \emph {et~al.}(2013)\citenamefont
  {Ahlgren}, \citenamefont {Ohlsson},\ and\ \citenamefont
  {Zhou}}]{Ahlgren:2013wba}%
  \BibitemOpen
  \bibfield  {author} {\bibinfo {author} {\bibfnamefont {B.}~\bibnamefont
  {Ahlgren}}, \bibinfo {author} {\bibfnamefont {T.}~\bibnamefont {Ohlsson}}, \
  and\ \bibinfo {author} {\bibfnamefont {S.}~\bibnamefont {Zhou}},\ }\href
  {\doibase 10.1103/PhysRevLett.111.199001} {\bibfield  {journal} {\bibinfo
  {journal} {Phys.Rev.Lett.}\ }\textbf {\bibinfo {volume} {111}},\ \bibinfo
  {pages} {199001} (\bibinfo {year} {2013})},\ \Eprint
  {http://arxiv.org/abs/1309.0991} {arXiv:1309.0991 [hep-ph]} \BibitemShut
  {NoStop}%
\bibitem [{\citenamefont {Cyr-Racine}\ and\ \citenamefont
  {Sigurdson}(2014)}]{Cyr-Racine:2013jua}%
  \BibitemOpen
  \bibfield  {author} {\bibinfo {author} {\bibfnamefont {F.-Y.}\ \bibnamefont
  {Cyr-Racine}}\ and\ \bibinfo {author} {\bibfnamefont {K.}~\bibnamefont
  {Sigurdson}},\ }\href {\doibase 10.1103/PhysRevD.90.123533} {\bibfield
  {journal} {\bibinfo  {journal} {Phys.Rev.}\ }\textbf {\bibinfo {volume}
  {D90}},\ \bibinfo {pages} {123533} (\bibinfo {year} {2014})},\ \Eprint
  {http://arxiv.org/abs/1306.1536} {arXiv:1306.1536 [astro-ph.CO]} \BibitemShut
  {NoStop}%
\bibitem [{\citenamefont {Archidiacono}\ and\ \citenamefont
  {Hannestad}(2014)}]{Archidiacono:2013dua}%
  \BibitemOpen
  \bibfield  {author} {\bibinfo {author} {\bibfnamefont {M.}~\bibnamefont
  {Archidiacono}}\ and\ \bibinfo {author} {\bibfnamefont {S.}~\bibnamefont
  {Hannestad}},\ }\href {\doibase 10.1088/1475-7516/2014/07/046} {\bibfield
  {journal} {\bibinfo  {journal} {JCAP}\ }\textbf {\bibinfo {volume} {1407}},\
  \bibinfo {pages} {046} (\bibinfo {year} {2014})},\ \Eprint
  {http://arxiv.org/abs/1311.3873} {arXiv:1311.3873 [astro-ph.CO]} \BibitemShut
  {NoStop}%
\bibitem [{\citenamefont {Dobrescu}\ and\ \citenamefont
  {Frugiuele}(2014)}]{Dobrescu:2014fca}%
  \BibitemOpen
  \bibfield  {author} {\bibinfo {author} {\bibfnamefont {B.~A.}\ \bibnamefont
  {Dobrescu}}\ and\ \bibinfo {author} {\bibfnamefont {C.}~\bibnamefont
  {Frugiuele}},\ }\href {\doibase 10.1103/PhysRevLett.113.061801} {\bibfield
  {journal} {\bibinfo  {journal} {Phys.Rev.Lett.}\ }\textbf {\bibinfo {volume}
  {113}},\ \bibinfo {pages} {061801} (\bibinfo {year} {2014})},\ \Eprint
  {http://arxiv.org/abs/1404.3947} {arXiv:1404.3947 [hep-ph]} \BibitemShut
  {NoStop}%
\bibitem [{\citenamefont {Heeck}\ and\ \citenamefont
  {Rodejohann}(2011)}]{Heeck:2011wj}%
  \BibitemOpen
  \bibfield  {author} {\bibinfo {author} {\bibfnamefont {J.}~\bibnamefont
  {Heeck}}\ and\ \bibinfo {author} {\bibfnamefont {W.}~\bibnamefont
  {Rodejohann}},\ }\href {\doibase 10.1103/PhysRevD.84.075007} {\bibfield
  {journal} {\bibinfo  {journal} {Phys. Rev.}\ }\textbf {\bibinfo {volume}
  {D84}},\ \bibinfo {pages} {075007} (\bibinfo {year} {2011})},\ \Eprint
  {http://arxiv.org/abs/1107.5238} {arXiv:1107.5238 [hep-ph]} \BibitemShut
  {NoStop}%
\bibitem [{\citenamefont {Blum}\ \emph {et~al.}(2014)\citenamefont {Blum},
  \citenamefont {Hook},\ and\ \citenamefont {Murase}}]{Blum:2014ewa}%
  \BibitemOpen
  \bibfield  {author} {\bibinfo {author} {\bibfnamefont {K.}~\bibnamefont
  {Blum}}, \bibinfo {author} {\bibfnamefont {A.}~\bibnamefont {Hook}}, \ and\
  \bibinfo {author} {\bibfnamefont {K.}~\bibnamefont {Murase}},\ }\href@noop {}
  {\  (\bibinfo {year} {2014})},\ \Eprint {http://arxiv.org/abs/1408.3799}
  {arXiv:1408.3799 [hep-ph]} \BibitemShut {NoStop}%
\bibitem [{\citenamefont {Forero}\ \emph {et~al.}(2014)\citenamefont {Forero},
  \citenamefont {Tortola},\ and\ \citenamefont {Valle}}]{Forero:2014bxa}%
  \BibitemOpen
  \bibfield  {author} {\bibinfo {author} {\bibfnamefont {D.}~\bibnamefont
  {Forero}}, \bibinfo {author} {\bibfnamefont {M.}~\bibnamefont {Tortola}}, \
  and\ \bibinfo {author} {\bibfnamefont {J.}~\bibnamefont {Valle}},\ }\href
  {\doibase 10.1103/PhysRevD.90.093006} {\bibfield  {journal} {\bibinfo
  {journal} {Phys.Rev.}\ }\textbf {\bibinfo {volume} {D90}},\ \bibinfo {pages}
  {093006} (\bibinfo {year} {2014})},\ \Eprint {http://arxiv.org/abs/1405.7540}
  {arXiv:1405.7540 [hep-ph]} \BibitemShut {NoStop}%
\bibitem [{\citenamefont {Ade}\ \emph {et~al.}(2015)\citenamefont {Ade} \emph
  {et~al.}}]{Ade:2015xua}%
  \BibitemOpen
  \bibfield  {author} {\bibinfo {author} {\bibfnamefont {P.}~\bibnamefont
  {Ade}} \emph {et~al.} (\bibinfo {collaboration} {Planck}),\ }\href@noop {} {\
   (\bibinfo {year} {2015})},\ \Eprint {http://arxiv.org/abs/1502.01589}
  {arXiv:1502.01589 [astro-ph.CO]} \BibitemShut {NoStop}%
\bibitem [{\citenamefont {Learned}\ and\ \citenamefont
  {Pakvasa}(1995)}]{Learned:1994wg}%
  \BibitemOpen
  \bibfield  {author} {\bibinfo {author} {\bibfnamefont {J.~G.}\ \bibnamefont
  {Learned}}\ and\ \bibinfo {author} {\bibfnamefont {S.}~\bibnamefont
  {Pakvasa}},\ }\href {\doibase 10.1016/0927-6505(94)00043-3} {\bibfield
  {journal} {\bibinfo  {journal} {Astropart.Phys.}\ }\textbf {\bibinfo {volume}
  {3}},\ \bibinfo {pages} {267} (\bibinfo {year} {1995})},\ \Eprint
  {http://arxiv.org/abs/hep-ph/9405296} {arXiv:hep-ph/9405296 [hep-ph]}
  \BibitemShut {NoStop}%
\bibitem [{\citenamefont {Athar}\ \emph {et~al.}(2000)\citenamefont {Athar},
  \citenamefont {Parente},\ and\ \citenamefont {Zas}}]{Athar:2000rx}%
  \BibitemOpen
  \bibfield  {author} {\bibinfo {author} {\bibfnamefont {H.}~\bibnamefont
  {Athar}}, \bibinfo {author} {\bibfnamefont {G.}~\bibnamefont {Parente}}, \
  and\ \bibinfo {author} {\bibfnamefont {E.}~\bibnamefont {Zas}},\ }\href
  {\doibase 10.1103/PhysRevD.62.093010} {\bibfield  {journal} {\bibinfo
  {journal} {Phys.Rev.}\ }\textbf {\bibinfo {volume} {D62}},\ \bibinfo {pages}
  {093010} (\bibinfo {year} {2000})},\ \Eprint
  {http://arxiv.org/abs/hep-ph/0006123} {arXiv:hep-ph/0006123 [hep-ph]}
  \BibitemShut {NoStop}%
\bibitem [{\citenamefont {Yuksel}\ \emph {et~al.}(2008)\citenamefont {Yuksel},
  \citenamefont {Kistler}, \citenamefont {Beacom},\ and\ \citenamefont
  {Hopkins}}]{Yuksel:2008cu}%
  \BibitemOpen
  \bibfield  {author} {\bibinfo {author} {\bibfnamefont {H.}~\bibnamefont
  {Yuksel}}, \bibinfo {author} {\bibfnamefont {M.~D.}\ \bibnamefont {Kistler}},
  \bibinfo {author} {\bibfnamefont {J.~F.}\ \bibnamefont {Beacom}}, \ and\
  \bibinfo {author} {\bibfnamefont {A.~M.}\ \bibnamefont {Hopkins}},\ }\href
  {\doibase 10.1086/591449} {\bibfield  {journal} {\bibinfo  {journal}
  {Astrophys.J.}\ }\textbf {\bibinfo {volume} {683}},\ \bibinfo {pages} {L5}
  (\bibinfo {year} {2008})},\ \Eprint {http://arxiv.org/abs/0804.4008}
  {arXiv:0804.4008 [astro-ph]} \BibitemShut {NoStop}%
\end{thebibliography}%

\end{document}